%% file: neurips_2022.tex
\documentclass{article}

\usepackage[preprint, nonatbib]{neurips_2022}

\usepackage[utf8]{inputenc} 
\usepackage[T1]{fontenc}    
\usepackage{url}            
\usepackage{booktabs}       
\usepackage{amsfonts}       
\usepackage{nicefrac}       
\usepackage{microtype}      
\usepackage[toc,page]{appendix}

\usepackage{siunitx}
\let\svqty\qty
\usepackage{physics}
\let\qty\svqty

\input{topmatter}

\usepackage{comment}
\usepackage{caption, subcaption}
\usepackage{tikz}
\tikzset{>=latex}
\usepackage{pgfplots}
\usepgfplotslibrary{fillbetween}

\usepackage[symbol*]{footmisc}

\DefineFNsymbolsTM{otherfnsymbols}{%
  \textdagger    \dagger
  \textdaggerdbl \ddagger
  \textasteriskcentered *
  \textbardbl    \|%
  \textparagraph \mathparagraph
}%

\setfnsymbol{otherfnsymbols}

\title{Quantum Neural Architecture Search\\with Quantum Circuits Metric\\and Bayesian Optimization}

\author{
  Trong Duong\thanks{Equal contribution} \\
  KAIST\\
  \texttt{trongduong@kaist.ac.kr} \\
  \And
  Sang T. Truong$\textsuperscript{$\dagger$}$  \\
  Stanford University \\
  \texttt{sttruong@cs.stanford.edu} \\
  \And
  Minh Tam$\textsuperscript{$\dagger$}$ \\
  Aalto University \\
  \texttt{minh.phamnguyen@aalto.fi} \\
  \And
  Bao Bach$\textsuperscript{$\dagger$}$ \\
  HCMUT \\
  \texttt{bao.bachbbace12@hcmut.edu.vn} \\
  \And
  Ju-Young Ryu\\
  KAIST \\
  \texttt{jyryu98@kaist.ac.kr} \\
  \And
  June-Koo Kevin Rhee\\
  KAIST \\
  \texttt{rhee.jk@kaist.edu} \\
}
\begin{document}
\maketitle
\begin{abstract}
    Quantum neural networks are promising for a wide range of applications in the Noisy Intermediate-Scale Quantum era. As such, there is an increasing demand for automatic quantum neural architecture search. We tackle this challenge by designing a quantum circuits metric for Bayesian optimization with Gaussian process. To this goal, we propose a new quantum gates distance that characterizes the gates' action over every quantum state and provide a theoretical perspective on its geometrical properties. Our approach significantly outperforms the benchmark on three empirical quantum machine learning problems including training a quantum generative adversarial network, solving combinatorial optimization in the MaxCut problem, and simulating quantum Fourier transform. Our method can be extended to characterize behaviors of various quantum machine learning models.
\end{abstract}


\section{Introduction}
There has been hope that complex problems will be solved efficiently on universal fault-tolerant quantum computers. However, a useful fault-tolerant quantum computer would require a sustainably large number of high-fidelity logical qubits and gate operations. Recent advances in the field are making quantum devices with a few hundred physical qubits and limited error correction into reality. Even though it is still far from an ideal quantum computer, those near-term devices might have some uses in certain optimization problems. Generally one has to find the minimal value of an objective function that usually involves a Hamiltonian, which could be of quantum origin or be mapped onto from a classical objective value. For example, Maximum Cut (MaxCut) is an NP-complete combinatorial optimization problem \cite{garey1974some} involves finding a partition $\{A,B\}$ of a given graph such that it maximizes sum of edge weights for edges connecting $A$ and $B$. We assign a binary value $x_i$ to every node corresponding to its class. The classical objective function is $C(x) = \sum_{i,j=1}^n w_{ij}x_i(1-x_j)$, where $w_{ij}$ is the edge weight. By mapping $x_i \mapsto \frac{1}{2}(1-Z_i)$, where $Z_i$ is the Pauli $Z$ operator with matrix representation $\text{diag} (1,-1)$, the objective can be translated into a Hamiltonian
\begin{equation}
    H = \sum_{x\in \{0,1\}^n} C(x) \ket{x}\bra{x} \label{eq:maxcut-hamil}
\end{equation}
in a quantum approximate optimization algorithm. The problem becomes finding the extreme value $\max_{\ket{\psi} \in \hilbert} \bra{\psi} H \ket{\psi}$. This is also a general form of optimization problems to be solved using variational quantum models.

Since going through every element of the Hilbert space $\hilbert$ is impossible, we parametrize the search space by a Parametrized Quantum Circuit (PQC) or a Quantum Neural Network (QNN), expressed by unitary matrix $U(\theta)$ that applies on an initial state $\ket{0}$. Whether the codomain $\{U(\theta)\ket{0}\}$ is a good search space for a specific problem depends entirely on the QNN architecture. Prior work estimates how well the search space induced by a QNN approximates the Hilbert space~\cite{sim2019expressibility} and how good its learning capacity is from a statistical perspective~\cite{berezniuk2020scale}. The availability of Noisy Intermediate-Scale Quantum (NISQ) devices allows executing quantum circuits of low depth where QNN-based algorithms constitute a huge class. The question regarding how to specify a proper quantum neural architecture is addressed by utilizing certain underlying structure and symmetry of specific problems such as Quantum Approximate Optimization Algorithms~\cite{farhi2014} or Variational Quantum Eigensolver~\cite{peruzzo2014ves}. However, necessary domain knowledge is not always available in practice. This highlights the need for a method to automate the QNN design process. Some approaches resort to quantum characteristics to do cell-based search~\cite{ostaszewski2021structure}, adaptive wavefunction ansatzs~\cite{grimsley2019adaptive, tang2021qubit}, variable-structure circuit~\cite{bilkis2021semi}, and hardware-efficient models ~\cite{kandala2017hardware, du2022quantum}. Another line of work, on the other hand, borrows ideas from classical neural architecture search (NAS). Representatives for this class include genetic algorithms~\cite{rattew2019domain}, meta neural predictor ~\cite{zhang2021neural}, differentiable search~\cite{zhang2020differentiable}, and reinforcement learning~\cite{kuo2021quantum}. Many approaches from this class deploy a good final architecture that can even outperform expert's designs. However, those methods behave in a black-box manner and often provide little insight for the quantum view. 

In this work we aim to apply Bayesian Optimization (BO), a well-known method for NAS~\cite{kandasamy2018neural, ru2020interpretable}, to Quantum NAS (QNAS). General information and setting of QNN and BO will be introduced in Section~\ref{Sec:Preliminaries}. Applying BO to QNAS requires an appropriate covariance function in space of QNN. The covariance function, or \textit{kernel}, is a key component to the success of the BO problem. Our main contribution, which is presented in Section~\ref{Sec:Methods}, is an operationally meaningful distance measure between two (possibly parametrized) quantum operators or \textit{quantum gates}, followed by a distance metric in the space of QNN constructed using optimal transport. Experiments of QNAS for three specific problems with the kernel based on that metric are presented in Section~\ref{Sec:Exp}. The paper finalizes with Section~\ref{Sec:result&discuss} and Section~\ref{Sec:Con&Future} where we  present the result along with some discussions and future work.

\section{Preliminaries} \label{Sec:Preliminaries}

\paragraph{Quantum Neural Networks} A QNN can be realized by a quantum circuit that contains a sequence of fixed and parametrized unitary operators, so-called \textit{quantum gates} acting on its qubits $q_0,q_1,..., q_n$, where $n$ is the number of processing qubit. Usually it is restricted to a sequence of fixed gates $W_i$ and one-parameter gates $U_i(\theta_i)$
\begin{equation}
    U(\theta) =  U_L(\theta_L) W_L \dots U_2(\theta_i) W_2 U_1(\theta_1) W_1.
\end{equation}
A fixed gate can be the \textit{identity} operator, hence two types of gates are not necessarily alternating. For this work we consider the most common logical gates in quantum computing~\cite{nielsenChuang, Qiskit}:
\begin{align}
    \nonumber &W_i \in \{H, X, Y, Z; CX, CY, CZ\} \\
    &U_i \in \{R_X(\cdot),R_Y(\cdot),R_Z(\cdot), CR_X(\cdot),CR_Y(\cdot),CR_Z(\cdot);R_{XX}(\cdot),R_{YY}(\cdot),R_{ZZ}(\cdot)\}
\end{align}
The semicolons separate single-qubit gates in the first half and two-qubit gates in the second half. Natural objective functions for training QNN is usually the expectation value of a Hermitian measurement operator $\Mc$ evaluated at the final state $ U(\theta)\ket{\psi}$ obtained by applying the QNN on an initial state $\ket{\psi}$, denoted by $E(\ket{\psi}, \theta) = \bra{\psi} U(\theta)^\dagger \Mc U(\theta) \ket{\psi}$. Optimizing the objective value can be done with a classical optimization algorithm.

\textbf{Bayesian Optimization} The goal of Bayesian Optimization (BO) is to find the global optimum of a fixed but unknown function $f: \Xc \rightarrow \Rbb$, which is potentially expensive and noisy to evaluate. We assume that $f$ is drawn from a prior distribution $p(f)$. Starting with an arbitrary dataset $\Dc_0$, BO sequentially queries the function to update our belief about it with guidance from an acquisition function, which measures the increase in utility of having an additional data point $(x, y = f(x) + \epsilon)$~\cite{brochu2010tutorial}. A common choice to model $p(f)$ is through Gaussian Process (GP). A GP describes a collection of random variables with a joint normal distribution identified by a mean function $m: \Xc \rightarrow \Rbb$ and a positive definite kernel $k: \Xc \times \Xc \rightarrow \Rbb$~\cite{rasmussen2003gaussian}. The GP can be used as a surrogate model for a black-box function $f(x) \sim \mathcal{GP}(m(x), k(x,X))$, where $X = [x_1, \cdots, x_{|\Dc|}]^T, Y = [y_1, \cdots, y_{|\Dc|}]^T$ are the observations we have made so far. With Gaussian prior, the posterior predictive $p(y | x, \Dc)$ is a Gaussian with mean function $m(x, X)$ and covariance function $k(x, X)$:
\begin{align}
    m(x, X) &= K(x, X) K_X^{-1} (Y - m(X)) \notag \\
    k(x, X) &= K(x, x) - K(x,X) K_X^{-1} K(X,x) \notag \\
    y &\sim \Nc(m(x, X), k(x,X))
    \label{eq:GP}
\end{align}
where $K(A,B)$ is a matrix whose the $(i,j)$-th element is computed as $k(x_i, x_j)$ with $x_i \in A$ and $x_j \in B$, and $K_X = K(X, X) + \sigma^2 I$ is the covariance between all noisy observations. The hyperparameters of the model are parameters of the mean and covariance functions and can be learned by maximizing the marginal likelihood of the input data. In this paper we use the Expected Improvement (EI) acquisition function, which measures the expectation of improvement over the current maximum value
\begin{equation} \label{eq:EI}
    \text{EI}_t(x) = \mathbb{E}[(f(x) - f^*_{t-1})^+ | (x_i,y_i)_{i=1}^{t-1}],
\end{equation}
where $f^*_{t-1} = \max_{i\leq t-1} f(x_i)$ is the best value up to time $t$, and $g(x)^+ = \max(0, g(x))$.

\section{Methods} \label{Sec:Methods}

Three representations of a QNN in our work are gate-based circuit, directed acyclic network (DAG), and vector representation as shown in Fig.~\ref{fig:qnn-representations}. The circuit representation is used for optimizing the objective function $\max_{\theta} E(\ket{\psi},\theta)$. The DAG representation provides a graph view that reveals the topological structure of the QNN. The vector representation only serves as the placeholder of the QNN and will not be used for anything except for sampling steps in BO. So, it should not raise a concern for being numerically arbitrarily chosen.

\tikzset{eln/.style={midway, font = \scriptsize,circle,inner sep=2pt}}
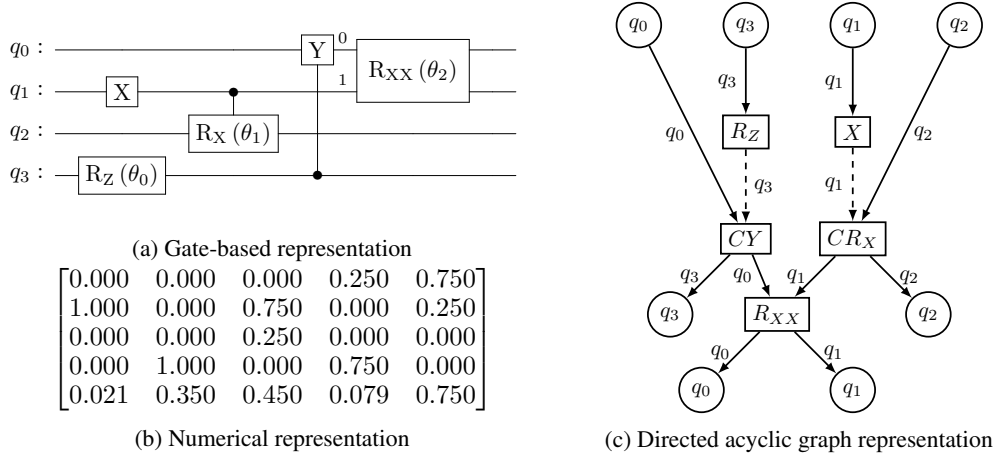
\begin{figure}[hbt!]
\centering
\begin{minipage}[c]{0.59\textwidth}
\begin{subfigure}{\textwidth}
    \includegraphics[width=0.9\textwidth]{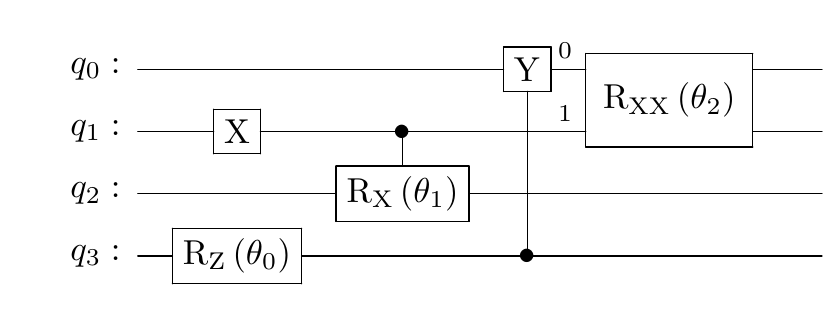}
    \caption{Gate-based representation}
    \label{fig:circ-repr}
\end{subfigure}
\begin{subfigure}{\textwidth}
    \centering
    $\begin{bmatrix}
   0.000 &   0.000 &   0.000 &   0.250 &   0.750 \\
   1.000 &   0.000 &   0.750 &   0.000 &   0.250 \\
   0.000 &   0.000 &   0.250 &   0.000 &   0.000 \\
   0.000 &   1.000 &   0.000 &   0.750 &   0.000 \\
   0.021 &   0.350 &   0.450 &   0.079 &   0.750 
    \end{bmatrix}$
    \caption{Numerical representation}
    \label{fig:vector-repr}
\end{subfigure}
\end{minipage}
\begin{minipage}[b][][b]{0.4\textwidth}
\begin{subfigure}{\textwidth}
    \centering
    \resizebox{0.9\textwidth}{!}{
    \begin{tikzpicture}[
            node distance={17mm}, thick, 
            qcircle/.style={draw, circle},
            qsquare/.style={draw},
        ]
        
        \node[qcircle]       (1)              {$q_0$}; 
        \node[qcircle]       (2) [right of=1] {$q_3$}; 
        \node[qcircle]       (3) [right of=2] {$q_1$};
        \node[qcircle]       (4) [right of=3] {$q_2$};
        
        \node[qsquare]       (5) [below of=2] {$R_Z$}; 
        \node[qsquare]       (6) [below of=3] {$X$};

        \node[qsquare]       (7) [below of=5] {$CY$}; 
        \node[qsquare]       (8) [below of=6] {$CR_X$};
        
        \node[qcircle]       (9) [below left of=7] {$q_3$}; 
        \node[qsquare]       (10)[below left of=8] {$R_{XX}$};
        \node[qcircle]       (11)[below right of=8] {$q_2$};

        \node[qcircle]       (12)[below left of=10] {$q_0$};
        \node[qcircle]       (13)[below right of=10] {$q_1$};
        
        \draw[->] (1) -- (7)    node[midway,left]{$q_0$};
        \draw[->] (2) -- (5)    node[midway,left]{$q_3$};
        \draw[->] (3) -- (6)    node[midway,left]{$q_1$};
        \draw[->] (4) -- (8)    node[midway,right]{$q_2$};
        \draw[->] (5) -- (7)    [dashed] node[midway,right]  {$q_3$};
        \draw[->] (6) -- (8)    [dashed] node[midway,left] {$q_1$};
        \draw[->] (7) -- (9)    node[midway,left]{$q_3$};
        \draw[->] (7) -- (10)   node[midway,left]{$q_0$};
        \draw[->] (8) -- (10)   node[midway,left]{$q_1$};
        \draw[->] (8) -- (11)   node[midway,right]{$q_2$};
        \draw[->] (10) -- (12)  node[midway,left]{$q_0$};
        \draw[->] (10) -- (13)  node[midway,right]{$q_1$};
    \end{tikzpicture}
    }
    \caption{Directed acyclic graph representation}
    \label{fig:dag-repr}
\end{subfigure}
\end{minipage}
  \caption{Three representations of a QNN. (a) The gate-based representation. (b) The DAG $\Gc=(\Lc,\Ec)$ with a set of gates $\Lc$ and directed connections between them $\Ec$. Note that each DAG node is sensitive to the order of incoming edges, which make the DAG and the gate-based circuit representations invertible. (c) The (matrixized) vector representation of the QNN. In general the matrix is of size $(n+1) \times N$, where $n$ is the number of qubits and $N$ the number of gates. The last row encodes the gate type, where every gate is given a \textit{representative number}, an evenly spaced value in the interval $(0,0.1)$ for each fixed gate and the interval $(0.1,1)$ for each parametrized gate. For every gate in the circuit, the row(s) of highest value/two highest values among the first $n$ rows in the corresponding column specifies the qubit the gate acts on. In the mapping from the matrix to the circuit, at each column, we choose the gate such that its representative number is closest to the value in the last row. Depending on its type, the row(s) with highest value(s) specifies qubit argument(s) of the gate. The mapping is not invertible; however it does not matter since the vector representations only stand as continuous placeholders for QNNs during sampling steps.}
  \label{fig:qnn-representations}
\end{figure}

\subsection{Distance Between Two Quantum Gates} \label{Subsec:GateDist}

\begin{figure}[hbt!] \centering
    \begin{subfigure}{0.24\textwidth} \centering
        \includegraphics[width=\textwidth]{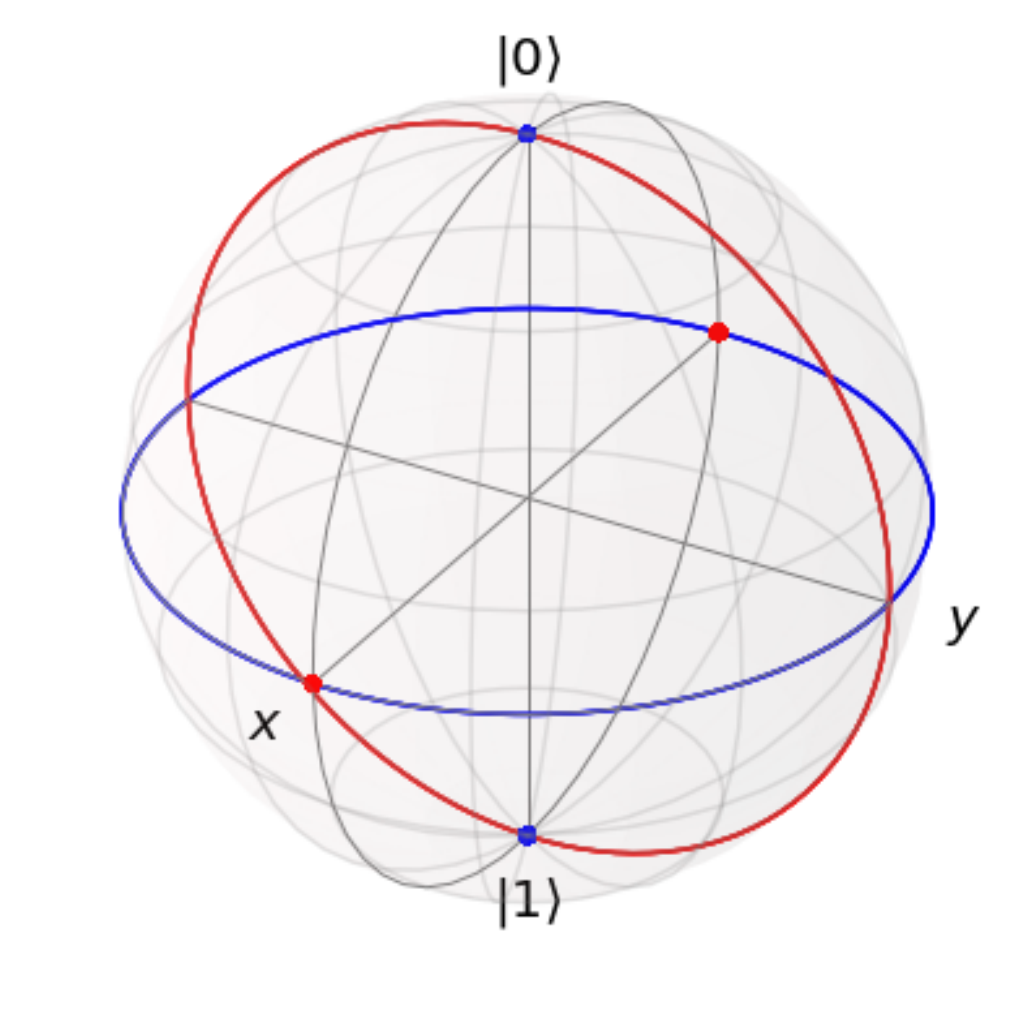}
        \caption{$R_Z$ and $R_X$}
    \end{subfigure}
    \begin{subfigure}{0.24\textwidth} \centering
        \includegraphics[width=\textwidth]{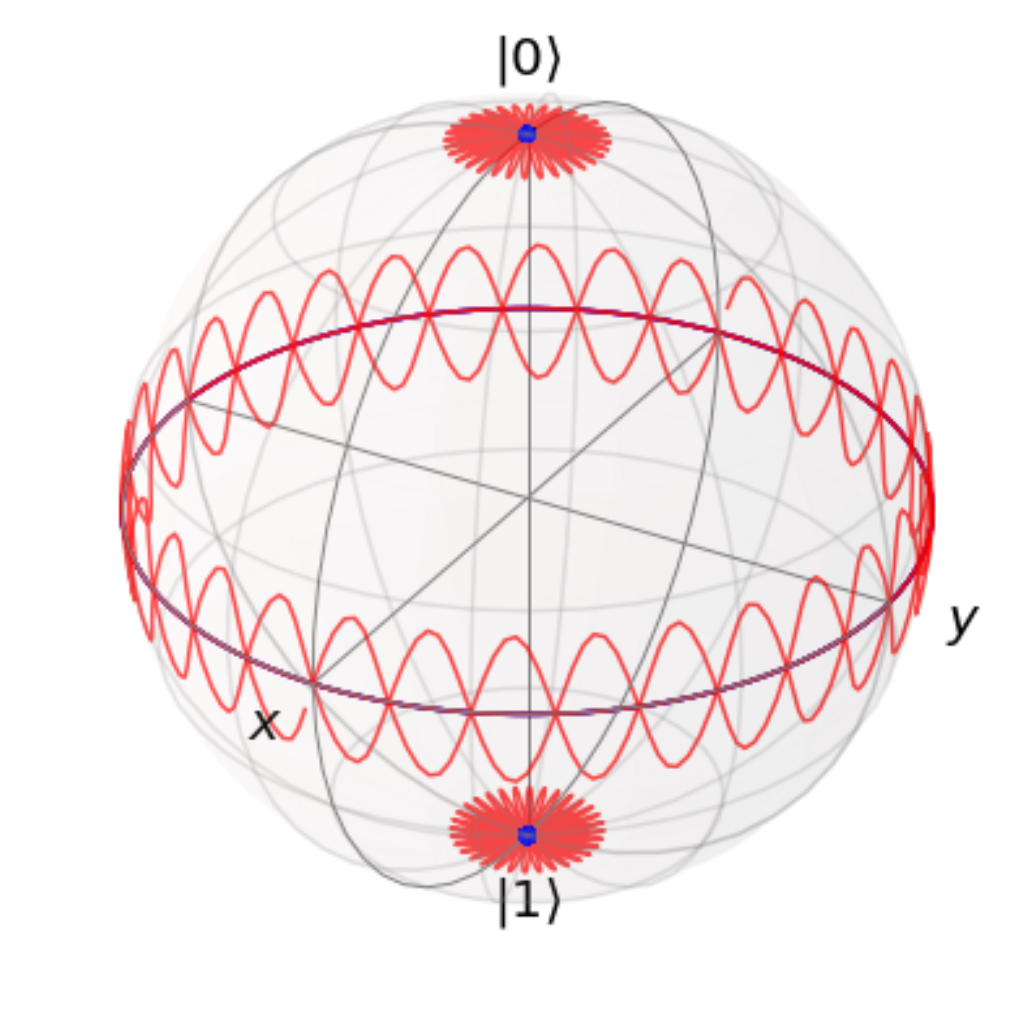}
        \caption{$R_Z$ and $R_Z^Y$}
    \end{subfigure}
    \begin{subfigure}{0.24\textwidth}\centering
        \includegraphics[width=\textwidth]{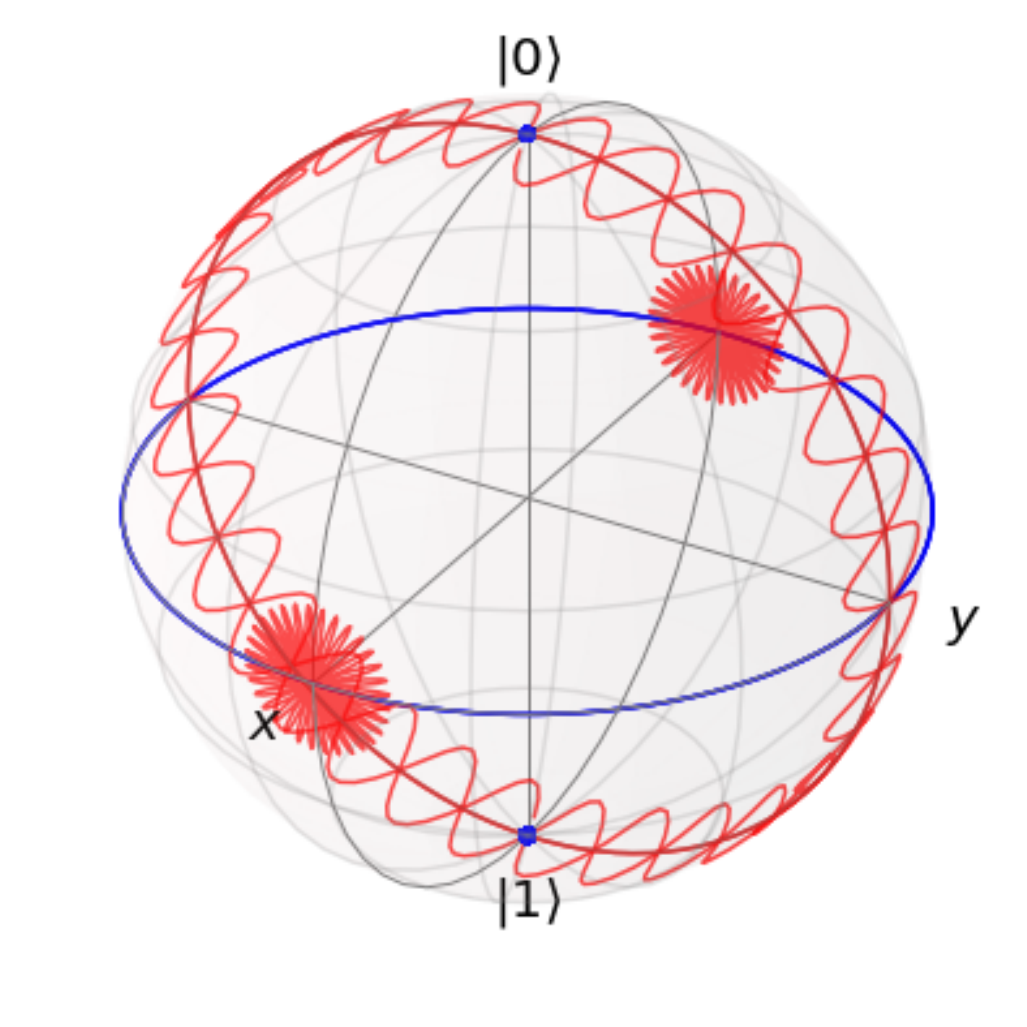}
        \caption{$R_Z$ and $R_X^Y$}
    \end{subfigure}   
    \begin{subfigure}{0.24\textwidth}\centering
        \includegraphics[width=\textwidth]{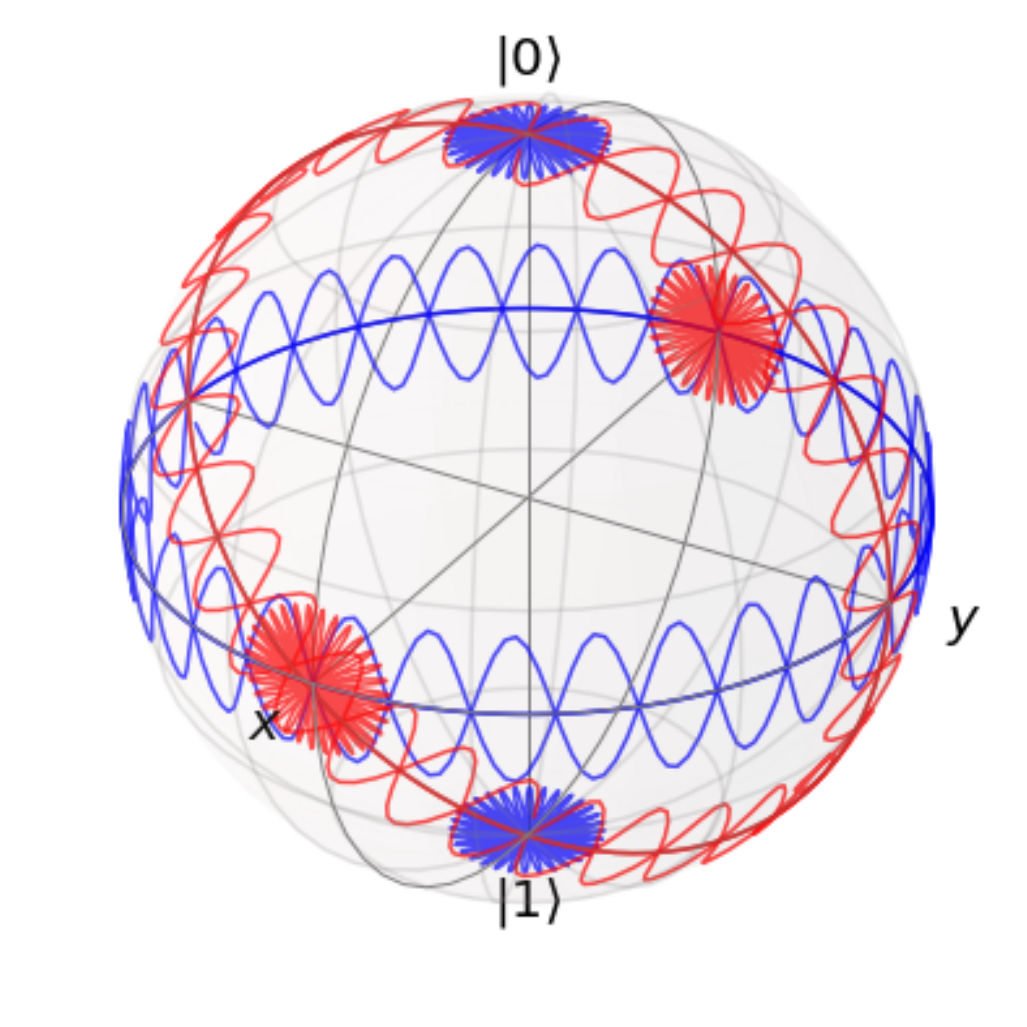}
        \caption{$R_Z^X$ and $R_X^Y$}
    \end{subfigure}       
    \caption{Orbits of single-qubit rotations (rotation direction in the subscript) and oscillatory rotations (oscillation direction in the superscript) when acting on $6$ anchor states $\ket{\pm x}, \ket{\pm y}, \ket{\pm z}$ of the one-qubit system. The blue/red line and dots indicate orbit of the first/second gate. \quad (a) $R_Z$ and $R_X$ trace out analogous circles and two clusters at the poles, $d_{\text{core}}(R_Z,R_X) > 0, d_{\text{shape}}(R_Z, R_X) = 0$. \quad (b) Both gates apply rotations around the $z$-axis, $d_{\text{core}}(R_Z,R_Z^Y) \approx 0$. \quad (c) $R_X^Y$ only has a small deviation from a circle $d_{\text{core}}(R_Z, R_X^Y) \approx d_{\text{core}}(R_Z,R_X)$,  $d_{\text{shape}}(R_Z,R_X^Y) \approx 0$. \quad (d) Two oscillatory rotations $R_Z^X$ and $R_X^Y$ trace out similar curves, $d_{\text{core}}(R_Z^X,R_X^Y) \approx d_{\text{core}}(R_Z,R_X), d_{\text{shape}}(R_Z^X,R_X^Y) = 0$.}
    \label{fig:gate-dist}
\end{figure}

We propose a similarity measure between two (possibly parametrized) quantum gates. Although the two unitary operations induced by the gates can be different, their effect might follow similar patterns throughout the Hilbert space. The difference between the exact effects will be captured by \textit{core distance} $d_\txtcore$, while the difference between their patterns by \text{shape distance} $d_\txtshape$. For example, one-qubit rotation gates are the most common gates in use; applying any rotation gate on a state with many angles always generates a circular orbit, so the shape distance between two rotations is always zero. Yet they revolve around different axes, which results in a positive core distance. Intuition about those two types of distance is illustrated in Fig.~\ref{fig:gate-dist}. We define the distance between two quantum gates to be the average between those two types of distance:

\begin{equation}
    d_{\text{gate}}(U^1, U^2) = \frac{d_{\text{core}}(U^1, U^2) + d_{\text{shape}}(U^1, U^2)}{2}
\end{equation}

\subsubsection{Core Distance}
Since every unitary operator is generated by some fixed Hermitian operator $H$ such that $U = e^{iHt}$, we can compare two unitary operators by their generating, assumably fixed, Hermitians. Since it is the solution to $\frac{dU}{dt} = iHU$ and $U$ decides the evolution of quantum states, $H$ governs the evolution's direction over time. For example, $R_X(\cdot)$ and $R_Y(\cdot)$ are rotations about the x- and y-axis. Their corresponding Hermitian operators are Pauli $X$ and $Y$. Hence it is natural to define a distance of $R_X$ and $R_Y$ using those Hermitian operators. Generally, we propose the difference between two generating Hermitian operators to be the \textit{core distance} between two quantum gates. For a quantum gate $U$, fixed or parametrized, there exists a unique scalar $t > 0$ and a unique Hermitian operator $H$ such that $\|H\|_* = 1$ and $U = e^{iHt}$. ($\| \cdot \|_*$ denote the nuclear norm; for Hermitian operators, the norm equals to the sum of absolute value of eigenvalues.) The core distance between the two unitary operators is given by 
\begin{equation}
d_{\text{core}}(U^1, U^2) = \frac{\|H^1 - H^2\|_*}{2}
\end{equation}
Since we only consider one-qubit and two-qubit gates, the unitary operator of the entire system of higher dimension is the tensor product of the gate's unitary operator with identity, e.g. $U = U_{\text{gate}} \otimes I$. The core distance is then affected by the qubits two gates apply on, i.e. in general, $d_\txtcore (U^1_{\text{gate}} \otimes I, U^2_{\text{gate}} \otimes I)$ is distinct to $d_\txtcore ( U^1_{\text{gate}} \otimes I, I \otimes U^2_{\text{gate}} )$.

\subsubsection{Shape Distance}
The shape distance represents the difference between orbital traces of two gates. For example, consider a one-qubit system which is represented by a Bloch sphere, the gate $R_X(\theta)$ traces out a circle around the $x$-axis. The orbit also has a circular shape for the gate $R_Y(\theta)$, except that the circle is around the $y$-axis. The shape distance should recognize their congruent shapes, i.e. $d_\txtshape(R_X,R_Y) = 0$. This shape comparison approach only makes sense when both gates in the pair are parametrized; in fact, every gate in the set of parametrized gates we choose for our work is a type of rotation parametrized by a single angle $\theta$. At the beginning, ignore the possible parameters, then two unitary operators $U^1$ and $U^2$ are the same if $|\bra{\psi}U^{2 \dagger} U^1\ket{\psi} |^2 = 1$ for every quantum state $\ket{\psi}$ in the Hilbert space $\Hc$, i.e.
\begin{align}
    \int_{\Hc} \left(1 - | \bra{\psi}U^{2 \dagger} U^1\ket{\psi} |^2 \right) d\mu(\psi) = 0,
\end{align}
where the integral is taken with the Haar measure~\cite{meckes2019random}. Simply speaking, it is the uniform distribution over unitary operators $U$, hence also the uniform distribution over quantum states $\ket{\psi} \equiv U\ket{0}$. To generalize the notion to parametrized quantum gates, think of $U^1(\cdot)\ket{\psi}$ and $U^2(\cdot) \ket{\psi}$ as two functions mapping $\theta$ to quantum states. The two gates are said to have the same shape when their codomains obtained as $\theta$ varies have the same shape up to a unitary transformation. This allows us to define the shape distance to be the minimal disparity induced by any unitary mapping. That is, two quantum gates are said to trace out the same orbit when for every quantum state $\ket{\psi} \in \hilbert$, there exists a quantum state $\ket{\phi}$ and a unitary operator $V$ such that
\begin{align}
    \left| \bra{\phi} U^{2\dagger}(\theta) V U^1(\theta)  \ket{\psi}\right|^2 = 1 \quad \forall \theta
\end{align}
The shape distance should consider the described fidelity over the Hilbert space, e.g.
\begin{align}
    \int_{\hilbert} \int_{\Theta} \left| \bra{\phi} U^{2\dagger}(\theta) V U^1(\theta)  \ket{\psi}\right|^2 d\theta d\mu(\psi)
\end{align}
We use a collection of quantum states $\{\ket{\psi_k}\}_{k=1}^K$ to approximate the integral. Members of the collection must represent the Hilbert space in some sense. We decide to use a collection called \textit{Mutually Unbiased Bases} (MUB) that contains $K = d(d+1)$ members for a qubit system of dimension $d=2^n$. A key property of MUB is that its members spread out evenly in the Hilbert space, where two distinct members are either orthonormal or have a fidelity of $\left|\braket{\psi_k}{\psi_{k'}}\right|^2 = 1/d$ \cite{durt2010mutually}. Moreover, the MUB also forms a complex projective 2-design that reduces an integral involving any unitary operator $U$ over every quantum state with the Haar measure to a finite sum ~\cite{klappenecker2005mutually}
\begin{align} \label{eq:complex-projective-design}
\int_{\hilbert} \left| \bra{\psi} U \ket{\psi} \right|^2 d\mu(\psi) = \frac{1}{K} \sum_{k=1}^K \left| \bra{\psi_k} U \ket{\psi_k} \right|^2
\end{align}
See Appendix~\ref{appendix:MUB} for further details about how to find that projective 2-design and its use in reduction of integral. We believe the symmetry of a MUB suffices to capture most of the information about the orbit traced out by acting a usual parametrized quantum gate on quantum states across the Hilbert space. Therefore we attempt to define the shape distance motivated by the following optimization problem
\begin{align}
    \nonumber & \min_{\substack{V, \phi_k}} \frac{1}{KT} \sum_{k,t} \left( 1 - \left| \bra{\phi_k} U^{2\dagger}(\theta_t) V U^1(\theta_t)\ket{\psi_k} \right|^2 \right) \\
    =& \min_{\substack{V, M}} \frac{1}{KT} \sum_{k,t} \left( 1 - \left| \bra{e_k}M^\dagger U^{2\dagger}(\theta_t) V U^1(\theta_t)\ket{\psi_k} \right|^2 \right) \label{eq:fidelity-form},
\end{align}
where $M \in \mathbb{C}^{d \times K}$ contains the (normalized) vector representation of every $\ket{\phi_k}$ in its columns, and $\{\ket{e_k}\}$ forms the standard basis of $\mathbb{C}^K$. However, we could not find a good way to optimize the $\| \cdot \|^2$ in \eqref{eq:fidelity-form}. Instead we define the shape distance with the root of fidelity
\begin{align}
\nonumber d_\txtshape(U^1, U^2) =& \min_{\substack{V, \phi_k}} \frac{1}{KT} \sum_{k,t} \left( 1 - \left| \bra{\phi_k} U^{2\dagger}(\theta_t) V U^1(\theta_t)\ket{\psi_k} \right| \right) \\
=& \min_{\substack{V, M}} \frac{1}{KT} \sum_{k,t} \left( 1 - \left| \bra{e_k}M^\dagger U^{2\dagger}(\theta_t) V U^1(\theta_t)\ket{\psi_k} \right| \right) \label{eq:root-fidelity-form}
\end{align}
Using the fact that $\min_{\alpha} \|e^{i\alpha} \ket{\psi} - \ket{\phi}\|^2 = 2 - 2|\braket{\phi}{\psi}|$, where the minimum is achieved at $\alpha = - \measuredangle \braket{\psi}{\phi}$ and $\measuredangle$ denotes the angle $\theta$ in the Euler's representation of a complex number $re^{i\theta}$, we can rewrite the problem with Euclidean distance. The distance, which is called MINT-OREO (\underline{M}aximum \underline{Int}egral of \underline{Or}bital \underline{E}mbedding \underline{O}verlaps), is implemented in the following form
\begin{align}
    d_\txtshape(U^1, U^2) =& \frac{1}{2KT} \min_{V,M,\alpha} \sum_{k,t} \|e^{i\alpha_{k,t}} VU^1(\theta_t)\ket{\psi_k} - U^2(\theta_t)M\ket{e_k} \|^2 \label{eq:shape-dist-final}
\end{align}
It is easy to see that $d_\txtshape(U^1, U^2) = 0$ when both gates are fixed. We assign $d_\txtshape(U^1, U^2) = \infty$ when a gate is fixed and the other is parametrized. If both gates are of parametrized type, we solve the optimization \eqref{eq:shape-dist-final} by coordinate descent for $\alpha,V,\alpha,M,\dots$ until convergence, where each update is optimal with respect to one variable. Finding the optimal unitary $V$ is known as \textit{Orthogonal Procrustes} with a closed-form solution~\cite{gower2004procrustes}. The optimal solution to $M$ can also be evaluated in closed form. We present details of the optimization and a discussion about the difficulty of optimizing the original sum-of-fidelity form in \eqref{eq:fidelity-form} in Appendix~\ref{apdx:mint-oreo}.

The MINT-OREO distance has some notable properties. First, it is invariant to the qubits $U^1$ and $U^2$ apply on, i.e. each of the two gates can apply on any qubit(s) of the system and the distance is unchanged. This is because $M$ can absorb the relative difference in the qubits being acted on by the two gates. Second, the distance is invariant to the class of parametrized gates. Let $A = \{R_X,R_Y,R_Z,R_{XX}, R_{YY}, R_{ZZ}\}, B = \{CR_X, CR_Y, CR_Z\}$ are the sets of rotations and controlled rotations. The intra-class distance is always $0$ and the inter-class distance is the same for every pair, that is $d(a,a') = 0, d(b,b')=0$, and $d(a,b) = d_{AB}$ for every $a,a' \in A, b,b' \in B$. We found out by experiments that the property holds for rotation in arbitrary directions. Third, the optimization converges in a few iterations and gives highly consistent results regardless of the initialization of $V$ and $M$. Also, it does not need to use many parameter samples; the result for $T=12$ only differs to that of $T=240$ by $\sim 10^{-3}$. Although we cannot prove it but the numerical result demonstrates that MINT-OREO distance forms a pseudo metric at least among quantum gates of rotation type (rotations and controlled rotations around arbitrary directions). This property still holds when we consider all of the fixed and parametrized gates.

In addition, the shape distance given by the finite sum is $0$ when $U_1(\theta)$ and $U_2(\theta)$ generate exactly the same shape. The result is stated formally in the following theorem, whose proof is also presented in Appendix~\ref{apdx:mint-oreo}.
\begin{theorem} \label{thm:integral-is-sum}
The shape distance given by the integral vanishes if and only if the shape distance given by a finite sum over anchor states vanishes. That is, for $T \geq 2$ distinct parameters $\theta_1, \dots, \theta_T$, there exist a unitary operator $V$ and quantum states $\ket{\phi}$ such that
\begin{equation}
    \sum_{t=1}^T \int_{\hilbert} \left(1 - \left| \bra{\phi} U^{2\dagger}(\theta_t) V U^1(\theta_t)  \ket{\psi}\right|\right) d\mu(\psi) = 0
\end{equation}
if and only if for the same $V$,
\begin{equation}
    \frac{1}{KT} \sum_{k,t} \left( 1 - \left| \bra{\phi_k} U^{2\dagger}(\theta_t) V U^1(\theta_t)\ket{\psi_k} \right| \right) = 0,
\end{equation}
where $\ket{\phi_k}$ is counterpart to $\ket{\psi_k}$ in the same way as $\ket{\phi}$ is counterpart to $\ket{\psi}$.
\end{theorem}

\subsection{Distance Between Two Quantum Neural Networks}
The similarity measure between any pair of gates can be generalized into a distance metric between two QNN architectures using optimal transport, whose application in classical NN architecture called OTMANN (Optimal Transport Metrics for Architectures of Neural Networks) was introduced by~\cite{kandasamy2018neural}. A simple interpretation of optimal transport program is to find a best way to move earth (dirt) piles from sources to destinations~\cite{rubner2000earth}. With the same analogy, the metric is defined as the minimum of a matching scheme that attempts to match the computation at the gates of one network to the gates of another, where penalties are given when there is any mismatch in the type of gates and topological difference between architectures. The minimum of those penalties is the OTMANN distance. Throughout this section, we view two QNNs as two DAGs $\mathcal{G}_1 = (\Lone, \mathcal{E}_1)$ and $\mathcal{G}_2 = (\Ltwo, \mathcal{E}_2)$ with $|\Lone| = n_1$ and $|\Ltwo| = n_2$ gates.

\begin{figure}[t] \centering
    \begin{subfigure}{0.32\linewidth} \centering
        \includegraphics[width=\textwidth]{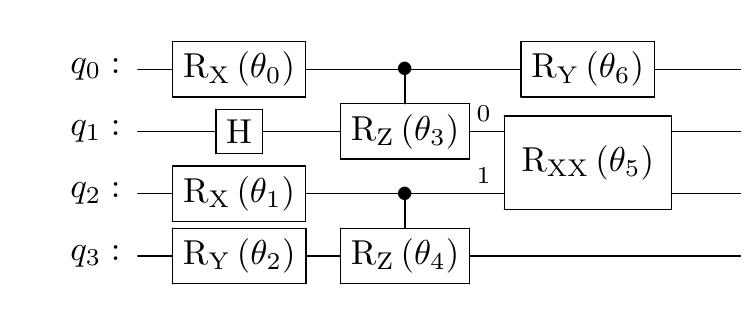}
        \caption{}
    \end{subfigure}
    \begin{subfigure}{0.32\linewidth} \centering
        \includegraphics[width=\textwidth]{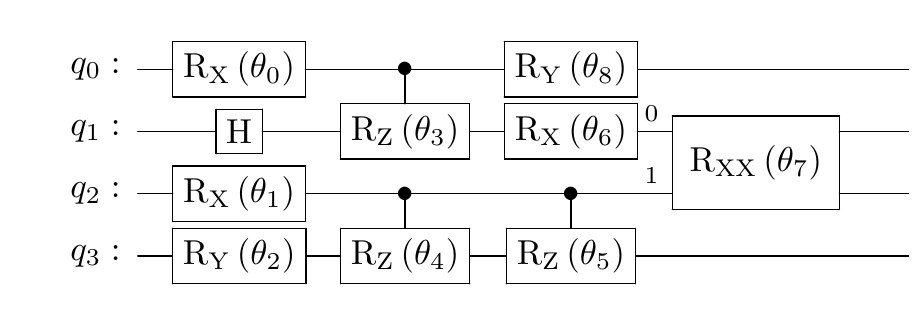}
        \caption{}
        \label{fig:qnn-dist-b}
    \end{subfigure}
    \begin{subfigure}{0.32\linewidth}\centering
        \includegraphics[width=\textwidth]{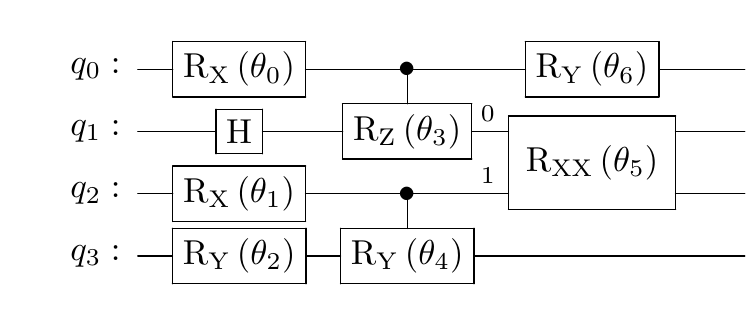}
        \caption{}
    \end{subfigure}   
    \caption{An illustration of some similar QNN architectures. The following are unnormalized distance $d$ and normalized distance $\bar d$ when $\nu= 0$: $d(a,b)=3.300, \bar d(a,b)=0.026; d(a,c)=5.303, \bar d(a,c)=0.042; d(b,c)=8.603, \bar d(b,c)=0.066$. Circuit (b) has a nonzero distance to (a) because of the extra $R_X$ gate on the $q_1$ wire. As every other gate in (b) has a perfect match in (a), that completely unmatched $R_X$ gate induces a distance of $3$ (its mass) + $0.1\times 3$ (extra mass gate $H$ gets due to the extra $R_X$). Two $CR_Z$ gates in $q_2$ and $q_3$ has the same effect and match to a single $CR_Z$ gate. The circuits (a) and (c) differ by a replacement of a $CR_Z$ gate by $CR_Y$ and has the distance equal to the gate distance of $0.3535$ multiplied by the common gate mass of $15$. In this case $d(b,c)=d(a,b) + d(a,c)$ because the differing components are independent.}
    \label{fig:qnn-dist}
\end{figure}

\paragraph{Gate Mass} Gate mass is a number representing the amount of computation a gate can perform. Motivated from the fact that a unitary matrix of size $p \times p$ has $p^2-1$ real degrees of freedom, we define the mass parametrized gates as $lm(u) = \text{param-dim}(u) \times (\text{unitary-dim}(u)^2 - 1)$. For example, single-qubit rotation gates have the mass of $3$, while controlled rotations and double rotations have the mass of $15$. The mass is even shared among gates of the same type if they are consecutive on the wire(s), e.g. two consecutive $CR_Z$ gates acting on $q_2$ and $q_3$ in Fig.~\ref{fig:qnn-dist-b} each has the mass of $15/2=7.5$.

Fixed gates, i.e. gates without parameters, on the other hand, have no value without the presence of parametrized gates. For instance, many quantum algorithms use fixed gates to initialize the quantum system into some favorable state such as a uniform superposition. The power of fixed gates in the circuit depends on the power of parametrized gates in the same circuit. Therefore, we assign them a fixed fraction of the layer mass of parametrized gates, i.e. for each fixed gate $u$, $lm(u) = \frac{\eta}{|\mathcal{DL}|} \sum_{s \in \mathcal{VL}} lm(s)$,
where $\mathcal{DL}$ and $\mathcal{VL}$ are the sets of fixed layers and parametrized layers, respectively, and $\eta$ is a fixed ratio. We choose $\eta = 0.1$ in our implementation. The total mass of an QNN is $tm(\mathcal{G}_i) = \sum_{u\in \mathcal{L}_i} lm(u)$.

\paragraph{Gate-type Mismatch Cost} This is the cost incurred when one attempts to match a gate $u_i \in \Lone$ to another gate $v_j \in \Ltwo$ of different type. It is natural to choose the gate distance defined in~\ref{Subsec:GateDist} for the mismatch cost. Hence the gate-type mismatch cost matrix $C_{\txtgtm} \in \Rbb^{n_1 \times n_2}$ is given by $(C_{\txtgtm})_{ij} = d_{\text{gate}} (u_i, v_j)$. This matrix can be immediately obtained from a list of precomputed gate-pair distances.

\paragraph{Structural Dissimilarity Cost} This term determines how different the relative position of two matched gates in their respective QNNs are. A small $(C_\txtstr)_{i,j}$ value means the gate $u_i \in \Lone$ and the gate $v_j \in \Ltwo$ are at structurally similar position. The cost matrix is computed by 
\begin{equation}
    (C_\txtstr)_{i,j} = \frac{1}{6n} \sum_{s \in \{\txtsp,\txtlp,\txtavg\}} \sum_{t \in \{\txtip,\txtop\}} \sum_{q=1}^n |\delta_t^{s,q}(i) - \delta_t^{s,q}(j)|,
\end{equation}
where ``sp'', ``lp'', ``avg'', ``ip'', ``op'' abbreviate for shortest path, longest path, average, input, and output. The value $\delta_t^{s,q}(\cdot)$ measures the shortest/longest/average path length from/to the input/output node of the $q$-th qubit in the respective DAG circuit. When there is no path between two nodes, we assign to it the longest path length in the entire DAG.

For instance, in Fig.~\ref{fig:dag-repr}, $\delta_{\txtsp/\txtlp/\txtavg}^{\txtip, 4}(CY) = 2$ because there is only one path from the input node of $q_3$ to the $CY$ node and it hops through two nodes including $CY$. On the other hand $\delta_{\txtsp/\txtlp/\txtavg}^{\txtop, 3}(CY) = 4$ because there is no path from the $CY$ node to $q_2$, so the longest path length of $4$ is assigned.

\paragraph{Optimal Transport Distance} 
Let $\braket{\cdot}{\cdot}$ denote the component-wise dot product and $\bfone_n$ an all-one vector. The optimal solution of the Kantorovich's formulation of Optimal Transport defines the distance between two QNNs. Let $\bar{n}_i = n_i+1; Z,C \in \Rbb^{\bar{n}_1\times \bar{n}_2}; y_1 \in \Rbb^{\bar{n}_1}, y_2 \in \Rbb^{\bar{n}_2}$ such that $\bfone^T_{\bar{n}_1}y_1 = \bfone^T_{\bar{n}_2}y_2$
\begin{equation}
    d(\mathcal{G}_1, \mathcal{G}_2) = \underset{Z}{\text{ minimize }} \langle Z,C \rangle \text{ s.t. } Z\geq 0, Z\bfone_{\bar{n}_2} = y_1, Z^T\bfone_{\bar{n}_1} = y_2
    \label{eq:otmann}
\end{equation}
In the formulation, $C$ is the matrix containing pairwise dissimilarity between component gates of the two circuits, or the ``ground distance'' in the optimal transport literature, and $Z=[z_{ij}]$ is the amount of mass of $u_i \in \Lone$ matched to $v_j \in \Ltwo$. We set $C = \begin{bmatrix} C_\txtgtm + \nu C_\txtstr & \bfone_{n_1} \\ \bfone^T_{n_2} & 0\end{bmatrix}$, $y_1 = \begin{bmatrix} \{lm(u)\}_{u \in \Lone}, tm(\mathcal{G}_2) \end{bmatrix}^T$, and $y_2 = \begin{bmatrix} \{lm(u)\}_{u \in \Ltwo}, tm(\mathcal{G}_1) \end{bmatrix}^T$. The last row and column in the cost matrix $C$ are dedicated to a made-up \textit{null gate} in each QNN. The weight of structural cost $\nu$ is a hyperparameter for the distance, usually set to $0.1$. Matching two gates is subject to gate-type mismatch and structural dissimilarity cost while matching a gate to the null gate induces to a cost of $1$, an upper bound for a finite $d_{\text{gate}}$ value. The null gate of a circuit is where a leftover mass of any gates in the other is collected when there is no possible match for it. 

The linear program can be solved efficiently by many solvers, and the resulting distance is a metric for quantum circuit architectures. The proof of the following theorem can be found in Appendix~\ref{apdx:OTmetric}. Also, Fig.~\ref{fig:qnn-dist} illustrates the distance concept through some similar-looking QNNs.

\begin{theorem}
The optimal solution $d(\mathcal{G}_1, \mathcal{G}_2)$ of \eqref{eq:otmann} forms a metric in the space of quantum circuit architectures when the weight of structural cost $\nu$ is positive. That is, for any quantum circuits $\mathcal{G}_1, \mathcal{G}_2, \mathcal{G}_3$ with described component gates, it holds that $d(\mathcal{G}_1, \mathcal{G}_2) \geq 0$, $d(\mathcal{G}_1, \mathcal{G}_2) = d(\mathcal{G}_2, \mathcal{G}_1)$, $d(\mathcal{G}_1,\mathcal{G}_3) \leq d(\mathcal{G}_1, \mathcal{G}_2) + d(\mathcal{G}_2, \mathcal{G}_3)$, and $d(\mathcal{G}_1,\mathcal{G}_2) = 0 \iff \mathcal{G}_1 = \mathcal{G}_2$.
\end{theorem}

\subsection{Bayesian Optimization}
The kernel we use for two QNNs $x$ and $x'$ in the Gaussian process is given by
\begin{equation}
    k(x,x') = \alpha e^{-\sum_{i}\beta_i d_i(x,x')} + \bar{\alpha} e^{-\sum_i\bar{\beta_i} \bar{d}_i(x,x')},
\end{equation}
where $d_i$ and $\bar{d}_i$ are unnormalized and normalized distances with some fixed $\nu_i$. It is important to remark that the two terms are not Gaussian and Laplace kernels although they have the same looking. A counterexample was given in \cite{peyre2019computational} to show the exponential of a negative squared optimal-transport distance does not necessarily induce a positive definite kernel. In our experiments, we, however, never encounter a situation where the covariance matrix has a negative eigenvalue whose absolute value exceeds a small jitter. We specify $\nu_i \in \{0.1, 0.2, 0.4, 0.8\}$ as default hyperparameters. Parameters of the model, including $\alpha, \bar{\alpha}, \beta_i, \bar{\beta}_i$ and the noise variance, are estimated via maximum likelihood.

Since a function on the space of architectures is not differentiable, it is common to use Evolutionary Algorithm (EA) to optimize acquisition functions \cite{kandasamy2018neural, white2019bananas, ma2019deep, ru2020interpretable}. We create a set of $N_{\text{mut}}$ samples at each generation by randomly mutating either the type or the acting wire(s) of a random gate in a circuit. Those with high acquisition function values will survive to the next generation with high probability. Although EA is not an efficient optimization algorithm for expensive functions, it works quite well for cheap evaluations of common acquisition functions. Further details are discussed in Appendix \ref{appdx:implementation}.

\section{Experiments} \label{Sec:Exp}
We demonstrate the QNN search for three problems: (i) simulate a Quantum Fourier Transform (QFT) (ii) solve MaxCut for graphs with random edge weights (iii) design a Quantum Generative Adversarial Net (QGAN), i.e. finding QNN architectures that optimize the corresponding black-box objective functions. 

First, QFT is a common subroutine in many quantum algorithms. Its unitary operator is $U_{\text{QFT}} = \frac{1}{\sqrt{d}} \sum_{s,t=0}^{d-1} \omega_d^{st}\ket{t}\bra{s}$ with system dimension $d = 2^n$ and phases $\omega_d^{st} = 2^{\pi i \frac{st}{d}}$. The QNN $U(\theta)$ learns to simulate the 2-qubit QFT by optimizing $\max_{\theta} \frac{1}{K} |\bra{\psi_k} U_{\text{QFT}}^\dagger U(\theta) \ket{\psi_k}|^2$ given the anchor states $\ket{\psi_k}$ as inputs to the network. 

Second, in the MaxCut problem, the QNN learns to separately maximize $M=10$ Hamiltonians corresponding to $M$ graphs with every edge having weight chosen uniformly random from $\{0,1,\dots,9\}$, i.e. $\frac{1}{M} \sum_{m=1}^M \max_{\theta_m} \frac{\bra{0} U^\dagger(\theta_m) H_m U(\theta_m) \ket{0}}{C_m}$, where $H_m$ is the Hamiltonian as described in \eqref{eq:maxcut-hamil} and $C_m$ the maximum cut value corresponding to the $m$-th graph. Note that a zero edge weight means the edge is neglected as if it does not exist. All the graphs in this experiment have $9$ nodes, which also needs 9 qubits. Finding the shape distance in the space of dimension $d=2^9$ is expensive, so we decide to use the shape distance for $d=2^4$ for this case. The shape distance of two gate types does not change much as the number of qubits increases, and $4$ qubits suffice for all relative positions possible when two quantum gates of single-qubit or two-qubit type are placed on the circuit.

Quantum Generative Adversarial Network (QGAN) is an adapted version of Generative Adversarial Network (GAN)~\cite{goodfellow2014generative} in which either or both the generator and the discriminator is realized by a QNN~\cite{dallaire2018,lloyd2018}. While the discriminator is trained to distinguish between true and synthesized data samples, the generator attempts to mimic the true data distribution by fooling the discriminator. In this experiment we train a QGAN that involves a quantum generator against a classical discriminator for learning and loading a probability distribution. In particular we use a 3-qubit QNN generator to generate samples following an even mixture of two normal distributions $\mathcal{N}(0.5,1^2)$ and $\mathcal{N}(3.5,0.5^2)$ truncated and discretized into bins $0,\dots,7$. The classical discriminator is a simple neural network with three layers as described in \cite{zoufal2019}. For the QGAN problem, we want to minimize the relative entropy, which is the KL divergence between the generated distribution $P$ and the true distribution $Q$ given by $D_{KL}(P||Q) = \sum_{x} P(x)\log \frac{P(x)}{Q(x)}$.

The initialization of all experiments starts by randomly sampling and training $5$ architectures. The implementation are made using Qiskit \cite{Qiskit} for quantum circuit design and BoTorch \cite{balandat2020botorch} for Bayesian Optimization framework. The setting for training those objectives is presented in Appendix \ref{appdx:implementation}.

\section{Results and Discussion} \label{Sec:result&discuss}

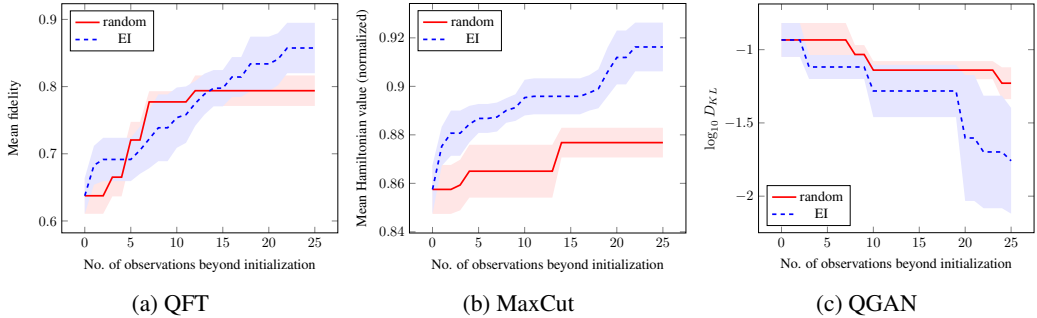
\begin{figure}[hbt!] \centering 
    \begin{subfigure}{0.325\linewidth} \centering
    \resizebox{\textwidth}{!}{
        \begin{tikzpicture}
        \begin{axis}[
            xlabel= No. of observations beyond initialization,
            ylabel= Mean fidelity,
            y label style={at={(axis description cs:0.05,.5)},rotate=0,anchor=south},
            legend pos=north west,
        ]
            \addplot [name path=lower, fill=none, draw=none, forget plot] table [
                x=X, y expr=\thisrow{Y} - \thisrow{STD}]{data/qft-random.txt};
            \addplot [name path=upper, fill=none, draw=none, forget plot] table [
                x=X, y expr=\thisrow{Y} + \thisrow{STD}]{data/qft-random.txt};
            \addplot[red!10, forget plot] fill between[of=lower and upper];
            \addplot [
                color=red,
                very thick
                ] table [x=X, y=Y]{data/qft-random.txt};
        
            \addplot [name path=lower, fill=none, draw=none, forget plot] table [
                x=X, y expr=\thisrow{Y} - \thisrow{STD}]{data/qft-EI.txt};
            \addplot [name path=upper, fill=none, draw=none, forget plot] table [
                x=X, y expr=\thisrow{Y} + \thisrow{STD}]{data/qft-EI.txt};
            \addplot[blue!10, forget plot] fill between[of=lower and upper];
            \addplot [
                color=blue,
                dashed,
                very thick
                ]  table [x=X, y=Y]{data/qft-EI.txt};
        \legend{random,EI}
        \end{axis}
        \end{tikzpicture}
        }
        \caption{QFT}
    \end{subfigure}
    \begin{subfigure}{0.325\linewidth} \centering
    \resizebox{\textwidth}{!}{
        \begin{tikzpicture}
        \begin{axis}[
            xlabel= No. of observations beyond initialization,
            ylabel= Mean Hamiltonian value (normalized),
            y label style={at={(axis description cs:0.05,.5)},rotate=0,anchor=south},
            legend pos=north west,
        ]
            \addplot [name path=lower, fill=none, draw=none, forget plot] table [
                x=X, y expr=\thisrow{Y} - \thisrow{STD}]{data/maxcut-random.txt};
            \addplot [name path=upper, fill=none, draw=none, forget plot] table [
                x=X, y expr=\thisrow{Y} + \thisrow{STD}]{data/maxcut-random.txt};
            \addplot[red!10, forget plot] fill between[of=lower and upper];
            \addplot [
                color=red,
                very thick
                ] table [x=X, y=Y]{data/maxcut-random.txt};
        
            \addplot [name path=lower, fill=none, draw=none, forget plot] table [
                x=X, y expr=\thisrow{Y} - \thisrow{STD}]{data/maxcut-EI.txt};
            \addplot [name path=upper, fill=none, draw=none, forget plot] table [
                x=X, y expr=\thisrow{Y} + \thisrow{STD}]{data/maxcut-EI.txt};
            \addplot[blue!10, forget plot] fill between[of=lower and upper];
            \addplot [
                color=blue,
                dashed,
                very thick
                ]  table [x=X, y=Y]{data/maxcut-EI.txt};
                
        \legend{random, EI}
        \end{axis}
        \end{tikzpicture}
        }
        \caption{MaxCut}
    \end{subfigure}
    \begin{subfigure}{0.325\linewidth}\centering
    \resizebox{\textwidth}{!}{
        \begin{tikzpicture}
        \begin{axis}[
            xlabel= No. of observations beyond initialization,
            ylabel= $\log_{10} D_{KL}$ ,
            y label style={at={(axis description cs:0.05,.5)},rotate=0,anchor=south},
            legend pos=south west,
        ]
            \addplot [name path=lower, fill=none, draw=none, forget plot] table [
                x=X, y expr=\thisrow{Y} - \thisrow{STD}]{data/qgan-random.txt};
            \addplot [name path=upper, fill=none, draw=none, forget plot] table [
                x=X, y expr=\thisrow{Y} + \thisrow{STD}]{data/qgan-random.txt};
            \addplot[red!10, forget plot] fill between[of=lower and upper];
            \addplot [
                color=red,
                very thick
                ] table [x=X, y=Y]{data/qgan-random.txt};
        
            \addplot [name path=lower, fill=none, draw=none, forget plot] table [
                x=X, y expr=\thisrow{Y} - \thisrow{STD}]{data/qgan-EI.txt};
            \addplot [name path=upper, fill=none, draw=none, forget plot] table [
                x=X, y expr=\thisrow{Y} + \thisrow{STD}]{data/qgan-EI.txt};
            \addplot[blue!10, forget plot] fill between[of=lower and upper];
            \addplot [
                color=blue,
                dashed,
                very thick
                ] table [x=X, y=Y]{data/qgan-EI.txt};
        \legend{random, EI}  
        \end{axis}
        \end{tikzpicture}
        }
        \caption{QGAN}
    \end{subfigure}
        
    \caption{Performance results. QFT: 2 qubits, 6 gates, 6 trials. MaxCut: 9 qubits, 5 gates, 6 trials. QGAN: 3 qubits, 12 gates, 3 trials. The y axes show the optimal objective values for learning problems. Confident band represents $\pm 1 \sigma$.}
    \label{fig:experiments}
\end{figure}

Fig.~\ref{fig:experiments} plots the best objective results for three optimization problems over time that compare the performance of our QNAS algorithm using Expected Improvement (``EI'') against random architecture sampling (``random''). We find that the QNAS algorithm performs consistently and significantly better than random sampling, common ansatzs, and problem-specific ansatzs, as presented in Table~\ref{tab:summary}. In the QFT experiment, the algorithm found some circuits with $6$ gates that simulate the QFT operator perfectly. They outperform common ansatzs with many more gates and parameters, and requires less gates (6 gates) than the most optimized QFT circuit (7 gates) in Qiskit using the same pool of component gates. In the MaxCut experiment, architectures found by BO perform better than the MaxCut ansatz and random architectures by a large margin. It might be because the ansatz is a trotterization of the Hamiltonian which, for a high-dimension system, requires a very large depth to converge; otherwise it lacks the flexibility for learning. This suggests we could use simpler circuits to find approximate solution to hard optimization problems. Finally, we found circuits with $12$ gates that has $D_{KL} \approx 10^{-3}$ in the QGAN experiment. The original paper use a circuit template with $24$ gates and longer training time to achieve the same precision \cite{zoufal2019}.

\begin{table}[hbt!] \centering
\makebox[\textwidth][c]{
\begin{tabularx}{\textwidth}{c|ccc|c|ccc}
    \toprule
     & {QFT} & {MaxCut} & {QGAN} & {} & {QFT} & {MaxCut} & {QGAN} \\ \midrule
    {\textbf{Bayes. Opt.}} & & & &
    {\textbf{19 az}.~\cite{sim2019expressibility}} & & & \\
    {EI} & \B 1 & \B 0.94 & \B 0.002 &
    {depth 1} & 0.98 & & 0.007 \\
    {random} & 0.88 & 0.90 & 0.065 & 
    {depth 2} & 1 & &  \\\midrule
    
    {\textbf{MaxCut az.}~\cite{farhi2014}} & & & &
    {\textbf{QGAN az.}~\cite{zoufal2019}}  & & & \\ 
    {depth 1} & & 0.746 & &
    {depth 1} & 0.7 & & 0.476\\
    {depth 2} & & 0.751 & &
    {depth 2} & 0.7 & & 0.277\\ 
    {depth 3} & & 0.762 & &
    {depth 3} & 0.7 & & 0.092\\ \bottomrule
\end{tabularx}
}
    \caption{Performance of QNN obtained by four ways on three experiments. Regarding methods, we report the best objective value over all searched architectures for BO and over 19 ansatz (az.) templates. As MaxCut and QGAN ansatzs are fixed, we report the average value after a number of training trials. A field is omitted when it is inconvenient or highly time-consuming to use the combination.}
    \label{tab:summary}
\end{table}

\section{Conclusion and Future Work} \label{Sec:Con&Future}
We presented a BO algorithm for quantum neural architecture search, which could find good architectures when the the desired output states are not very restrictive. The main contribution is a geometrically meaningful distance between quantum gates and the extension to distance between quantum circuit architectures using optimal transport. The gate distance and circuit distance could have independent usages beyond BO algorithms. A possible direction is to examine the relation between the gate/circuit distance and learning behaviors of QNNs. Another direction is to design the generalization of the core distance when the Hamiltonian is not fixed and the shape distance when the parameter spaces for $U^1$ and $U^2$ are distinct , which could provide a way to analyze data generated by quantum processes.


\bibliography{references}

\newpage

\appendix
\section{Mutually Unbiased Bases}
\subsection{Complex Projective 2-design}
Reducing the integral over the Haar measure to a simple sum of terms of the same form, as in \eqref{eq:complex-projective-design} is possible given the right representatives of the space. The collection of such representatives of the unitary space is called a quantum unitary design.

\begin{definition}
Let $P_{t,t}: \Uc(\Hc) \rightarrow \R$ be a polynomial with homogeneous degree at most $t$ in the entries of $W \in  \Uc(\Hc)$ and their complex conjugate. A finite set of unitary operators $\{W_k\}_{1}^K$ forms a unitary $t$-design when
\begin{equation}
    \frac{1}{K} \sum_{k=1}^K P_{t,t}(W_k) = \int_{\Uc(\Hc)}P_{t,t}(W) d\mu(W),
\end{equation}
where $\mu$ is the Haar measure, holds for all such polynomials $P_{t,t}$.

For any unitary $t$-design $\{W_k\}$ and any fixed pure state $\ket{\psi}$, the set $\{W_k \ket{\psi}\}$ is a complex projective $t$-design.
\end{definition}

Since the expression is exact, and the integrand $|\bra{\psi} U \ket{\psi}|^2 \equiv |\bra{0}W^\dagger U W\ket{0}|^2$ for some $W$ in \eqref{eq:complex-projective-design} is a polynomial of degree $2$ in entries of $W$ and $W^\dagger$, it can be computed given a unitary $2$-design or a complex projective design. It is known from quantum information that the Clifford group forms a unitary $3$-design that also makes it a $2$-design by definition. However, the size of the group grows exponentially with the dimension of the Hilbert space. Another option is Mutually Unbiased Bases (MUB), which is a complex projective $2$-design and only has $d(d+1)$ element vectors when $\Hc \subset \mathbb{C}^d$. An efficient construction of the MUB when the dimension is a power of a prime based on Galois field and Galois ring introduced by~\cite{klappenecker2003constructions} is briefly introduced.

\subsection{Construction of Mutually Unbiased Bases for Prime Power Dimension Space} \label{appendix:MUB}
Mutually unbiased bases in the Hilbert space $\mathbb{C}^d$ are orthonormal bases in which every two bases $\{\ket{e_1}, \dots, \ket{e_d}\}$ and $\{\ket{f_1}, \dots, \ket{f_d}\}$ satisfy
\begin{equation}
    | \braket{e_j}{f_k} |^2 = \frac{1}{d}, \quad \forall j,k
\end{equation}
When $d$ is a power of a prime number, a set of MUB contain exactly $d+1$ bases, and $d(d+1)$ basis vectors form a complex projective $2$-design~\cite{klappenecker2005mutually}.

\subsubsection[Odd prime power]{Odd prime power $d = p^n$}
Denote $\mathbb{F}_d$ a finite field with $d$ elements of odd characteristic $p$. The field trace map $\text{tr}: \mathbb{F}_d \rightarrow \mathbb{F}_p$ is defined by $\text{tr}(x) = x + x^p + x^{p^2} \dots + x^{p^{n-1}}$.

For every $a \in \mathbb{F}_d$, the set
\begin{align}
    B_a &= \{ \ket{v_{a,b}}: b\in \mathbb{F}_d \},
    \\
    \ket{v_{a,b}} &= d^{-1/2}(\omega_p^{\text{tr}(ax^2 + bx)})_{x \in \mathbb{F}_d} \in \mathbb{C}^d, \notag
\end{align}
where $\omega_p = \exp(2\pi i/p)$, is an orthonomal basis. Then the bases $B_a, a\in \mathbb{F}_q$ together with the standard basis form a set of $d+1$ mutually unbiased bases in $\mathbb{C}^d$.

\subsubsection[Even Prime Power]{Even Prime Power $d = 2^n$}
The method for odd prime power does not work in the case $p=2$, which is our interest for qubit systems. For a monic basic primitive polynomial $h(x)$ of degree $n$, the quotient ring $\text{GR}(4,n) = \mathbb{Z}_4[x] / \langle h(x) \rangle$ is the Galois ring of degree $n$ over $\mathbb{Z}_4$ (since all Galois ring constructed by different $h(x)$ are indeed isomorphic). There exists a nonzero element $\xi$ in $\text{GR}(4,n)$ with order $2^n - 1$. Any element $r \in \texttt{GR}(4,n)$ has a unique expression $r = a+2b$, where $a,b \in \Tc_n = \{0,1,\xi,\dots,\xi^{2^n-2}\}$. The \texttt{Teichmüller set}  $\Tc_n$ allows us to characterize elements of the ring in terms of $a,b$. The trace map $\text{tr}: \text{GR}(4,n) \rightarrow \mathbb{Z}_4$ is defined by $\text{tr}(x) = \sum_{k=0}^{n-1} \sigma^k (x)$, where $\sigma: \text{GR}(4,n) \rightarrow \text{GR}(4,n)$ defined by $\sigma(a+2b) = a^2 + 2b^2$ is called the Frobenius automorphism. Then the orthonormal bases $B_a, a\in \Tc_n$ together with the standard basis form a collection of MUB, where
\begin{equation}
\begin{aligned}
    B_a &= \{ \ket{v_{a,b}}: b\in \Tc_n\},
    \\
    \ket{v_{a,b}} &= d^{-1/2} \left( \exp \left( \frac{2\pi i}{4}\text{tr}((a+2b)x) \right) \right)_{x \in \Tc_n}
\end{aligned}
\end{equation}

\section{Maximum Integral of Orbital Embedding Overlaps (MINT-OREO) Distance}
\label{apdx:mint-oreo}
\subsection{Optimizing the shape distance}
The optimization problem \eqref{eq:shape-dist-final} can be written in two forms

\begin{align}
    &\min_{V,M} \sum_{k,t} 1 - \left| \bra{e_k}M^\dagger U^{2\dagger}(\theta_t) V U^1(\theta_t)\ket{\psi_k} \right| \notag \\
    =& \frac{1}{2} \min_{V,M,\alpha} \sum_{k,t} \|e^{i\alpha_{k,t}} VU^1(\theta_t)\ket{\psi_k} - U^2(\theta_t)M\ket{e_k} \|^2 \label{eq:fix-M} \\
    =& \frac{1}{2} \min_{V,M,\alpha} \sum_{k,t} \|e^{i\alpha_{k,t}} U^{2\dagger}(\theta_t)VU^1(\theta_t)\ket{\psi_k} - M\ket{e_k} \|^2 \label{eq:fix-V}
\end{align}

It can be solved by optimizing $\alpha,V,\alpha,M,\dots$ iteratively, where each iteration constitutes those four updates. We use the expression \eqref{eq:fix-M} to optimize for $\alpha, V$. Denote 

\begin{align}
    L_{1} & = [ e^{i\alpha_{1,1}}U^{1} (\theta_1) \ket{\psi_1} \dots  e^{i\alpha_{K,T}} U^{1} (\theta_T) \ket{\psi_K}], \\
    L_{2} & = [ U^{2} (\theta_1) M \ket{e_1} \dots  U^{2} (\theta_T) M \ket{e_K}]
\end{align}

the matrices in $\mathbb{C}^{d \times KT}$ that contains all outputs resulted from applying $U^1, U^2$. Note that $L_1, L_2$ contains $K$ groups of consecutive columns with $t=1,\cdots,T$ for each $k$. Let $L_1 L_2^\dagger =R\Sigma T^\dagger$ be a singular value decomposition (after update $\alpha$). The variables $\alpha$ and $V$ are updated by
\begin{align}
    \alpha_{k,t} \longleftarrow &  - \measuredangle \bra{ e_k} M^{\dagger}  U^{2\dagger}(\theta_t) V U^1(\theta_t)\ket{\psi_k} \\
    V \longleftarrow & R^\dagger T
\end{align}

The first rule results from the fact $\min_{\substack{\alpha}} \|e^{i\alpha}\ket{\phi} - \ket{\psi}\|^2 = 2-|\braket{\phi}{\psi}|$ at $\alpha = -\measuredangle \braket{\psi}{\phi}$, where $\measuredangle$ denotes the angle in the Euler's representation of a complex number. We can rewrite \eqref{eq:fix-M} as $\frac{1}{2} \|VL_1 - L_2\|_F^2$, whose minimization is equivalent to maximizing $\text{Re} \Tr [L_2^\dagger V L_1] = \text{Re} \Tr [\Sigma (T^\dagger V R)]$. It reaches the maximum of $\Tr(\Sigma)$ if $T^\dagger V R = I$ or $V = R^\dagger T$, which is the second rule.

We use the expression \eqref{eq:fix-V} to optimize for $\alpha,M$
\begin{align}
    \alpha_{k,t} \longleftarrow &  - \measuredangle \bra{ e_k} M^{\dagger}  U^{2,\dagger}(\theta_t) V U^1(\theta_t)\ket{\psi_k} \\
     M \longleftarrow & \begin{bmatrix} \hat M_1 \dots \hat M_k \dots \hat M_K \end{bmatrix}
\end{align}

Denote $S = [ \dots e^{i\alpha_{k,t}} U^{2\dagger}(\theta_t)VU^1(\theta_t)\ket{\psi_k} \dots]$ and $E = [\ket{e_1} (T \text{ times}), \dots, \ket{e_K} (T \text{ times})]$ with standard basis vectors $\{\ket{e_k}\}$ of  $\mathbb{C}^{K}$. After $\alpha$ is updated, the problem in \eqref{eq:fix-V} is equivalent to minimizing $\|S - ME\|_F^2$. We can optimize every column of $M$ independently. Let $M_k$ be its $k$-th column and $S_{k,1}, \cdots S_{k,T}$ be consecutive columns of $S$. The optimization for a column of $M$ can be formulated as
\begin{align}
    \min_{M_k} \sum_{t=1}^T \|M_k - S_{k,t}\|^2 \text{ s.t. } \|M_k\|^2 = 1
\end{align}
It is straightforward that the problem is equivalent to $\max_{M_k} \text{Re } (M_k^\dagger S_k)$ for $S_k = \sum_{t=1}^T S_{k,t}$. The optimal value for the column is $\hat M_k = \frac{S_k}{\|S_k\|}$ obtained from the fact that
\begin{align}
    \text{Re} \left(x^\dagger y\right) = 
    \begin{bmatrix}
    \text{Re } x^{T} &
    \text{Im } x^{T}
    \end{bmatrix}
    \begin{bmatrix}
    \text{Re } y \\
    \text{Im } y
    \end{bmatrix},
\end{align}
which is maximized when $x = \lambda y, \lambda \in \mathbb{R}$.
\subsection{Finite sum characterizes integral over the Hilbert space}
Now we prove Theorem.~\ref{thm:integral-is-sum}, which could be simplified to
\begin{equation}
    \frac{1}{T} \sum_{t=1}^T \int_{\hilbert} \left| \bra{\phi} U^{2\dagger}(\theta_t) V U^1(\theta_t)  \ket{\psi}\right| d\mu(\psi) = 1
\end{equation}
if and only if
\begin{equation}
    \left| \bra{\phi_k} U^{2\dagger}(\theta_t) V U^1(\theta_t)\ket{\psi_k} \right| = 1 \quad \forall k,t
\end{equation}
The forward direction is straightforward due to the continuity of $\ket{\phi}$. For the reverse statement, choose $t = 1$ as real phase references, i.e. $\ket{\phi_k} = U^{2\dagger}(\theta_1) V U^1(\theta_1) \ket{\psi_k}$ for every $k$. For every $t > 1$, $\ket{\phi_k} = e^{-i\alpha_{kt}} U^{2\dagger}(\theta_t) V U^1(\theta_t) \ket{\psi_k}$ for some real $\alpha_{kt}$. The identity $\braket{\phi_k}{\phi_k} = 1$ implies that
\begin{equation} \label{eq:behavior-on-basis}
    \bra{\psi_k} \left(U^{2\dagger}(\theta_1) V U^1(\theta_1)\right)^\dagger \left(U^{2\dagger}(\theta_t) V U^1(\theta_t)\right) \ket{\psi_k} = e^{i\alpha_{kt}}
\end{equation}
From the construction of MUB, the set of anchor states contains the standard basis $\{\ket{e_m}\}_{i=m}^{d}$. Without loss of generality, assume $\ket{\psi_m} = \ket{e_m}, m=1,\dots,d$. Since the action of a linear operator is completely defined with a basis, we can rewrite
\begin{equation}
    \left(U^{2\dagger}(\theta_1) V U^1(\theta_1)\right)^\dagger \left(U^{2\dagger}(\theta_t) V U^1(\theta_t)\right) \ket{e_m} = e^{i\alpha_{mt}} \ket{e_m}
\end{equation}
For ease of writing, denote $W_t = U^{2\dagger}(\theta_t) V U^1(\theta_t)$. Let $\ket{\phi} = W(\theta_1) \ket{\psi}$. The goal is to show $| \bra{\phi} W(\theta) \ket{\psi} | = 1$ for every $t=1,\dots,T$ and  $\ket{\psi} \in \hilbert$.

Let $\ket{\psi} = \sum_{m=1}^d c_m \ket{e_m}, \ket{\psi_k} = \sum_{m=1}^d c_{k,m} \ket{e_m}$ be normalized expressions in the standard basis. Then
\begin{align}
\bra{\phi} W_t \ket{\psi} = \bra{\psi} W_1^\dagger W_t \ket{\psi} = \left| \sum_{m=1}^d |c_m|^2 e^{i\alpha_{mt}}  \right|
\end{align}

For a fixed $t$, the phases $\{e^{i\alpha_{mt}}\}$ are not independent. From the first $d$ anchors we can already determine $e^{i\alpha_{mt}}$  for all $m$. However, the other $d^2$ anchors constrain $e^{i \alpha_{mt}}$ as
\begin{equation}
      \bra{\phi_k} W_t \ket{\psi_k} | = 1 \longrightarrow  \left| \sum_{m=1}^d |c_{k,m}|^2 e^{i \alpha_{mt}} \right| = 1 
\end{equation}
The above constraints are equivalent to $d^2$ linear equations in $d^2$ variables $e^{ -i \alpha_{mt}  } e^{i \alpha_{nt}}, m,n=1,\dots,d$. It has at most one solution, which occur when $e^{i \alpha_{mt}} = e^{i \alpha_{1t}}$. This completes the proof as $\left| \sum_{m=1}^d |c_m|^2 e^{i\alpha_{mt}}  \right| = \left| \sum_{m=1}^d |c_m|^2 e^{i\alpha_{1t}}\right| = 1$.

\subsection{Remark on \texorpdfstring{$L_2$}{L2} formulation of shape distance}
Back to the formulation for the shape distance, although we have not found an efficient way to optimize the sum-of-fidelity formulation 
\begin{equation}
\label{eq:sum-of-fidelity}
\max_{\substack{V, M }} \sum_{k=1}^{K} \sum_{t=1}^T  | \bra{ e_k } M^{\dagger} U^{2 \dagger} (\theta_t) V U^1 (\theta_t) \ket{\psi_k} |^2,
\end{equation}
it could have a meaning in quantum information beyond our initial motivation. To optimize V, we extend $\ket{\psi_k} \rightarrow \ket{\psi_k} \otimes \ket{k} \otimes 1, M \ket{e_k} \rightarrow (M\ket{e_k}) \otimes \ket{k} \otimes 1, U^i(\theta_t) \rightarrow U^i(\theta_t) \otimes 1 \otimes \ket{t}$ and $ V \rightarrow V \otimes I \otimes I $. Define $L_1 = [ U^{1} (\theta_1) \ket{\psi_1} \dots  U^{1} (\theta_T) \ket{\psi_K}]$ and $ L_2 = [ U^{2} (\theta_1) M \ket{e_1} \dots  U^{2} (\theta_T) M \ket{e_K}] $. We can rewrite \eqref{eq:sum-of-fidelity} using Frobenius norm
\begin{align}
    \nonumber & \sum_{k=1}^{K} \sum_{t=1}^T  | \bra{ e_k } M^{\dagger} U^{2 \dagger} (\theta_t) V U^1 (\theta_t) \ket{\psi_k} |^2 \\
    \nonumber =& \|L_2^\dagger (V \otimes I \otimes I) L_1\|_F^2 \\
    \nonumber =& \text{Tr} \left(W_2^\dagger \Sigma_2^\dagger R_2^\dagger (V \otimes I \otimes I) R_1 \Sigma_1 \Sigma_1^\dagger R_1^\dagger (V^\dagger \otimes I \otimes I) R_2 \Sigma_2 W_2     \right) \\
    =& \text{Tr} \left( \rho_2 (V \otimes I \otimes I) \rho_1 (V^\dagger \otimes I \otimes I) \right) 
\end{align}
One can view $\rho_i = R_i D_i R_i^\dagger$ as a density matrix consisting of a probe system and the environment, where diagonal matrices $D_i$ that contain nonnegative eigenvalues of $L_i L_i^\dagger$. The optimization problem $\max_{\substack{V}} \text{Tr} \left( \rho_2 (V\otimes I \otimes I) \rho_1 (V^\dagger \otimes I \otimes I) \right)$ could be interpreted as finding the best unitary mapping $V$ of the system to map $\rho_1$ to $\rho_2$, both of which are quantum states of the system and the environment. 

To optimize M, we follow the same idea as above by extending our system as followed $ \ket{e_k} \rightarrow \ket{e_k} \otimes \ket{k}, U^{2}(\theta_t) V U^{1} (\theta_t) \ket{\psi_k} \to \Big( U^{2}(\theta_t) V U^{1} (\theta_t) \ket{\psi_k} \Big) \otimes \ket{k}$ and $M \rightarrow M \otimes I $. We define $L_2^{\prime} = [ \ket{e_1} \dots \ket{e_K}  ] $ and $L_{1}^{\prime} = [ U^{2}(\theta_1) V U^{1} (\theta_1) \ket{\psi_1} \dots U^{2}(\theta_T) V U^{1} (\theta_T) \ket{\psi_K}  ]$. Using Frobenius norm, the maximization problem over M can be casted as 
\begin{equation}
    \max_{\substack{M }} \sum_{k=1}^{K} \sum_{t=1}^T  | \bra{ e_k } M^{\dagger} U^{2 \dagger} (\theta_t) V U^1 (\theta_t) \ket{\psi_k}| = \max_{\substack{M}} \text{Tr} \left( \rho_2^{\prime} (M^{\dagger} \otimes I ) \rho_1^{\prime} (M \otimes I) \right), 
\end{equation}
This problem has a similar form as optimizing $V$ except $M$ is only constrained to have normalized columns.

\section{Optimal-Transport Distance as a Metric}
\label{apdx:OTmetric}
The optimal-transport distance is non-negative and symmetric straightforward from its definition and constraints. Now we alter the settings to prove the triangle inequality. Let $tm(\mathcal{G}_i), i=1,2,3$ be the total mass of $\mathcal{G}_i, i=1,2,3$. Because transporting masses between two null gates takes no cost, the solution to \eqref{eq:otmann} is the same with $y_i = \begin{bmatrix} \{lm(u)\}_{u \in \mathcal{L}_i}, \sum_{j\neq i} tm(\mathcal{G}_j) \end{bmatrix}^T$. This is to make the total mass $\sum_{i=1}^3 tm(\mathcal{G}_i)$ for all three pairs, hence without generality we prove the triangle inequality assuming the total mass is $1$. We follow the classic proof in \cite{rubner2000earth}. Let $\{z^{(12)}_{ij}\}$ be the optimal flow from $\mathcal{G}_1$ to $\mathcal{G}_2$ and $\{z^{(23)}_{jk}\}$ the optimal flow from $\mathcal{G}_2$ to $\mathcal{G}_3$, i.e. the solutions to the corresponding optimal transport problems. Let $s_{ijk}$ be the mass that moves from gate $u_i \in \Lone$ to gate $v_j \in \Ltwo$ and from $v_j \in \Ltwo$ to $w_k \in \mathcal{L}_3$. Define a flow $z^{(13)_{ik}}$ from $\mathcal{G}_1$ to $\mathcal{G}_3$ by
\begin{equation}
    z^{(13)}_{ik} = \sum_{j} s_{ijk}
\end{equation}
This flow is feasible since
\begin{align}
    \nonumber \sum_{i} z^{(13)}_{ik} = \sum_{i,j} s_{ijk} = \sum_{j} z^{(23)}_{jk} = lm(w_k), \\
    \sum_{k} z^{(13)}_{ik} = \sum_{j,k} s_{ijk} = \sum_{j} z^{(12)}_{ij} = lm(u_i)
\end{align}
where $lm(u_i)$ ($lm(w_k)$) is the gate mass if $u_i$ ($w_k$) is a quantum gate or $ tm(\mathcal{G}_2) +  tm(\mathcal{G}_3)$ ($tm(\mathcal{G}_1) +  tm(\mathcal{G}_3)$) if it is a null gate. Then
\begin{align}
    \nonumber d( \mathcal{G}_1, \mathcal{G}_3 ) &\leq \sum_{i,k} z^{(13)}_{ik} c_{u_i, w_k} \\
    \nonumber &\leq \sum_{i,j,k} s_{ijk} c_{u_i, w_k} \\
    &\leq \sum_{i,j,k} s_{ijk} (c_{u_i, v_j} + c_{v_j, w_k}) \\
    \nonumber &= \sum_{i,j} z^{(12)}_{ij} c_{u_i, v_j} + \sum_{j,k} z^{(23)}_{jk} c_{v_j, w_k} \\
    \nonumber &= d( \mathcal{G}_1, \mathcal{G}_2 ) + d( \mathcal{G}_2, \mathcal{G}_3 )
\end{align}
The inequality in the third line comes from the fact that the cost between gates, including null ones, is metric. Because the cost is a weighted sum of core distance, shape distance, structural dissimilarity cost, and non-assignment cost, it suffices to show each component bears similar properties. The core distance with nuclear norm is a metric. Numerical results of the shape distance show it is a pseudo-metric for our set of gates as remarked in Section~\ref{Sec:Methods}. Regarding the non-assignment cost, if there is at least a null gate among $u_i, v_j, w_k$, the costs that involve the null gate are either $0$ or $1$ which leads to the triangle inequality; otherwise the term for non-assignment cost disappears. The structural dissimilarity cost satisfies the triangle inequality because the following holds trivially for real numbers $\delta_t^{s,q}(\cdot)$:
\begin{equation}
    | \delta_t^{s,q}(i) - \delta_t^{s,q}(k)  | \leq  | \delta_t^{s,q}(i) - \delta_t^{s,q}(j)  | +  | \delta_t^{s,q}(j) - \delta_t^{s,q}(k)  | 
\end{equation}
The optimal-transport distance is therefore a metric in the strict sense. Two quantum circuits must be the same when they have zero distance. Every gate in one circuit then must have an exact counterpart gate in the other, where they have the same relative position to all of the qubit wires and are of the same gate type. This could only happen when the two circuits are identical.

\section{Illustration of Quantum Circuits Metric}
\begin{figure}[h!] \centering
    \begin{subfigure}{0.49\textwidth} \centering
        \includegraphics[width=\textwidth]{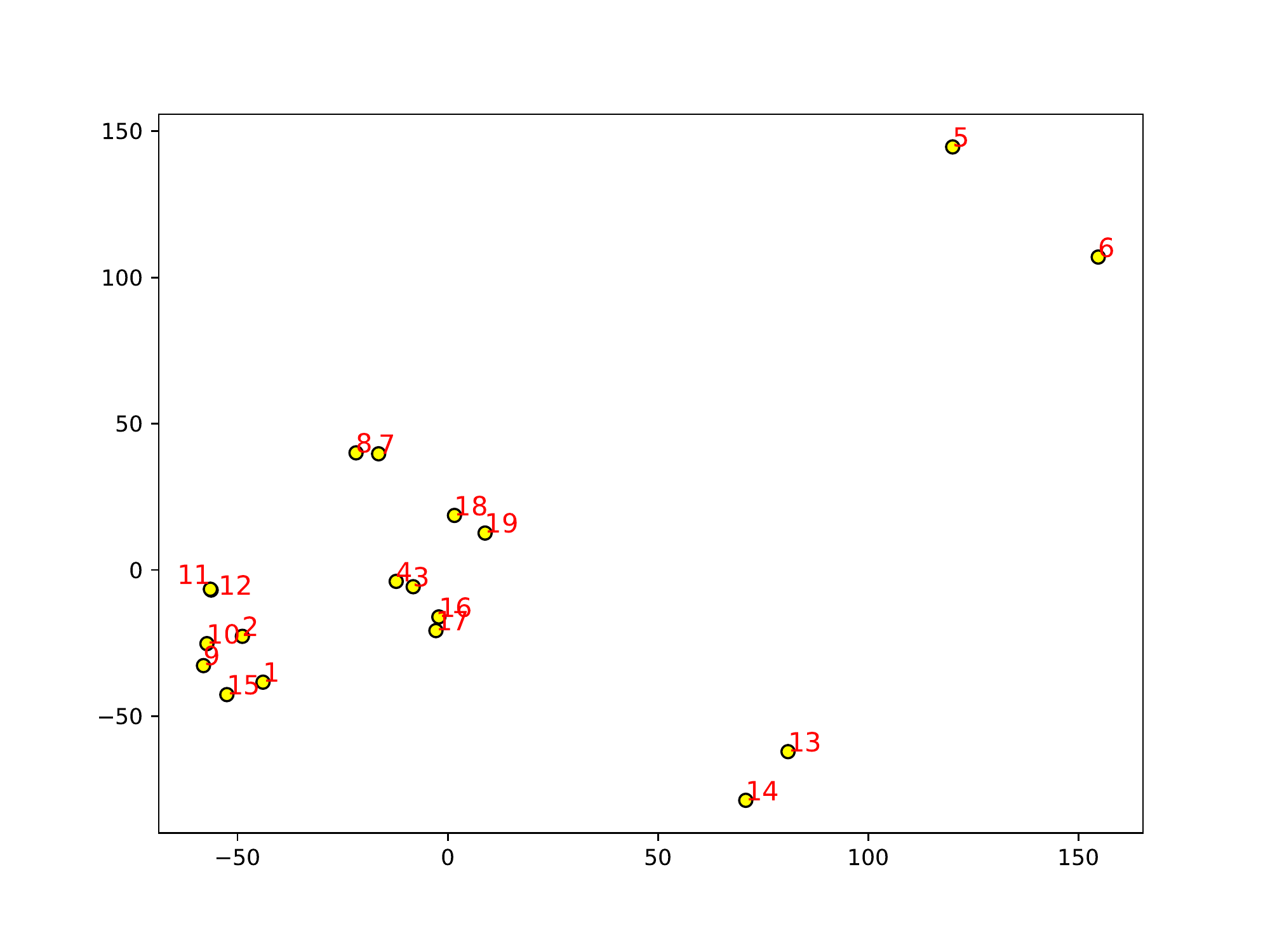}
        \caption{MDS: Circuit Distance}
    \end{subfigure}
    \begin{subfigure}{0.49\textwidth} \centering
        \includegraphics[width=\textwidth]{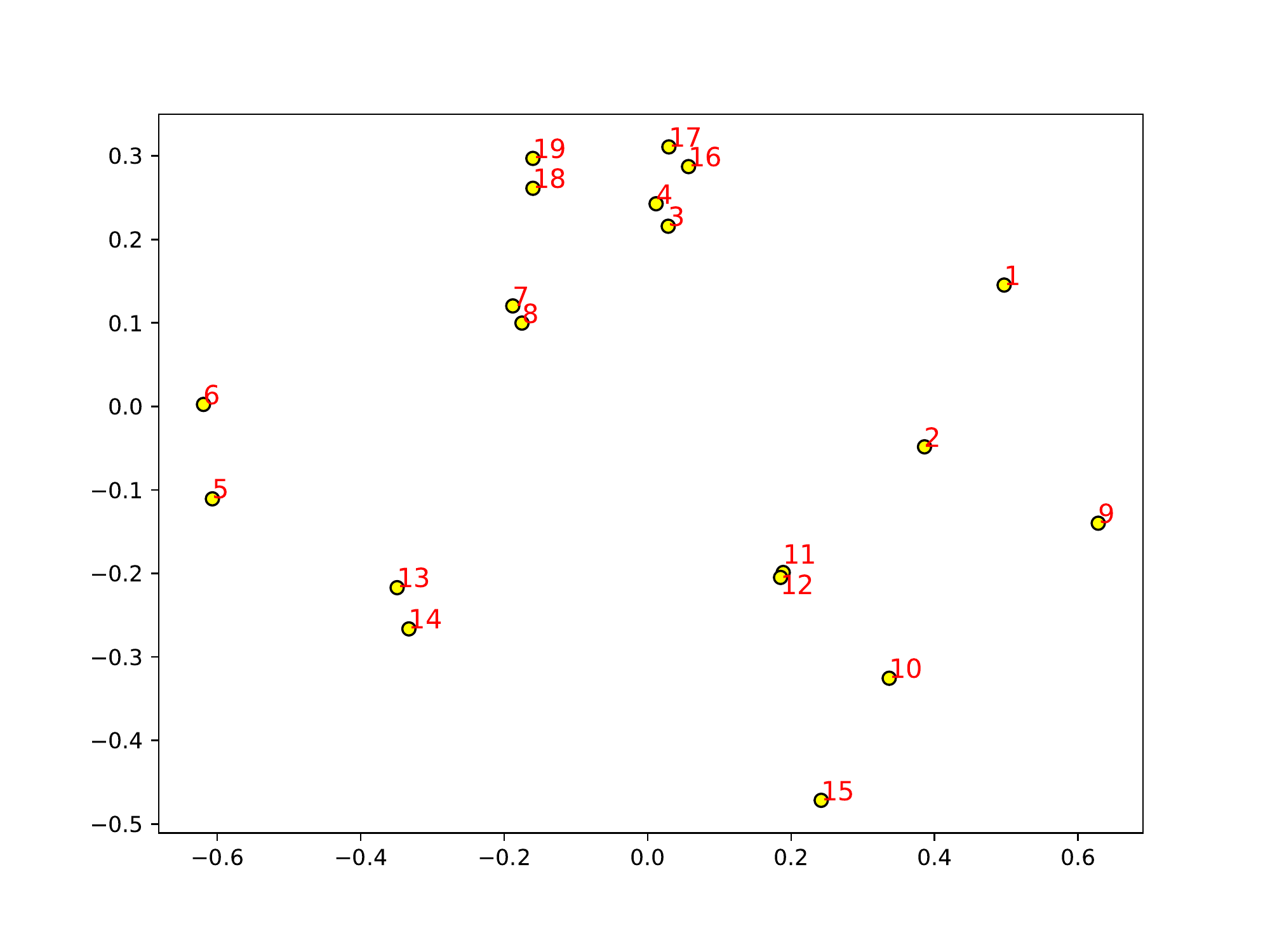}
        \caption{MDS: Normalized Circuit Distance}
    \end{subfigure}
    \caption{The two dimensional embedding from pairwise (normalized) distance with $\nu = 0.5$ of 19 circuit templates. The circuits are labeled with the same ID shown in Figure 2 in \cite{sim2019expressibility}. Circuits with notable similarity are close to each other such as (3,4), (5,6), (7,8), (11,12), (13,14), (16,17), (18,19). Remaining circuits (1,2,9,10,15) have fewer parametrized gates, hence small total mass and are placed within a neighborhood.}
    \label{fig:common-templates}
\end{figure}

First, we show the circuit metric as in \eqref{eq:otmann} agree with our intuition through a visualization of pairwise distances of some common circuit architectures using Multidimensional Scaling (MDS). The MDS assigns each architecture a point $(x_1,\dots,x_d) \in \mathbb{R}^d$ so that two points are close to each other if the distance between the two corresponding is small~\cite{mead1992review}. Fig.~\ref{fig:common-templates} shows a two-dimensional embedding of 19 circuit templates with 4 qubits surveyed in literature~\cite{sim2019expressibility}.

\begin{figure}[h!] \centering
    \begin{subfigure}{0.49\textwidth} \centering
        \includegraphics[width=\textwidth]{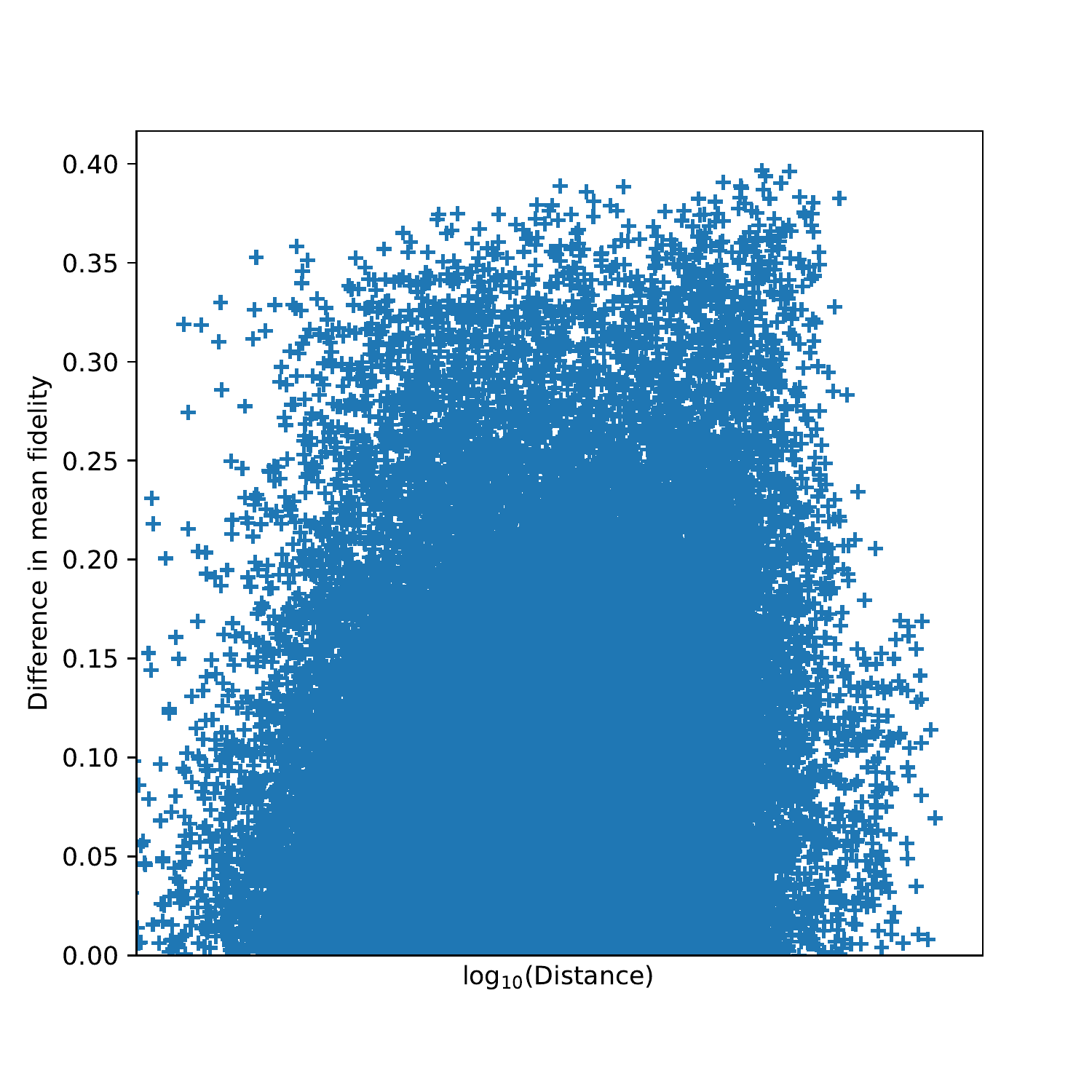}
    \end{subfigure}
    \begin{subfigure}{0.49\textwidth} \centering
        \includegraphics[width=\textwidth]{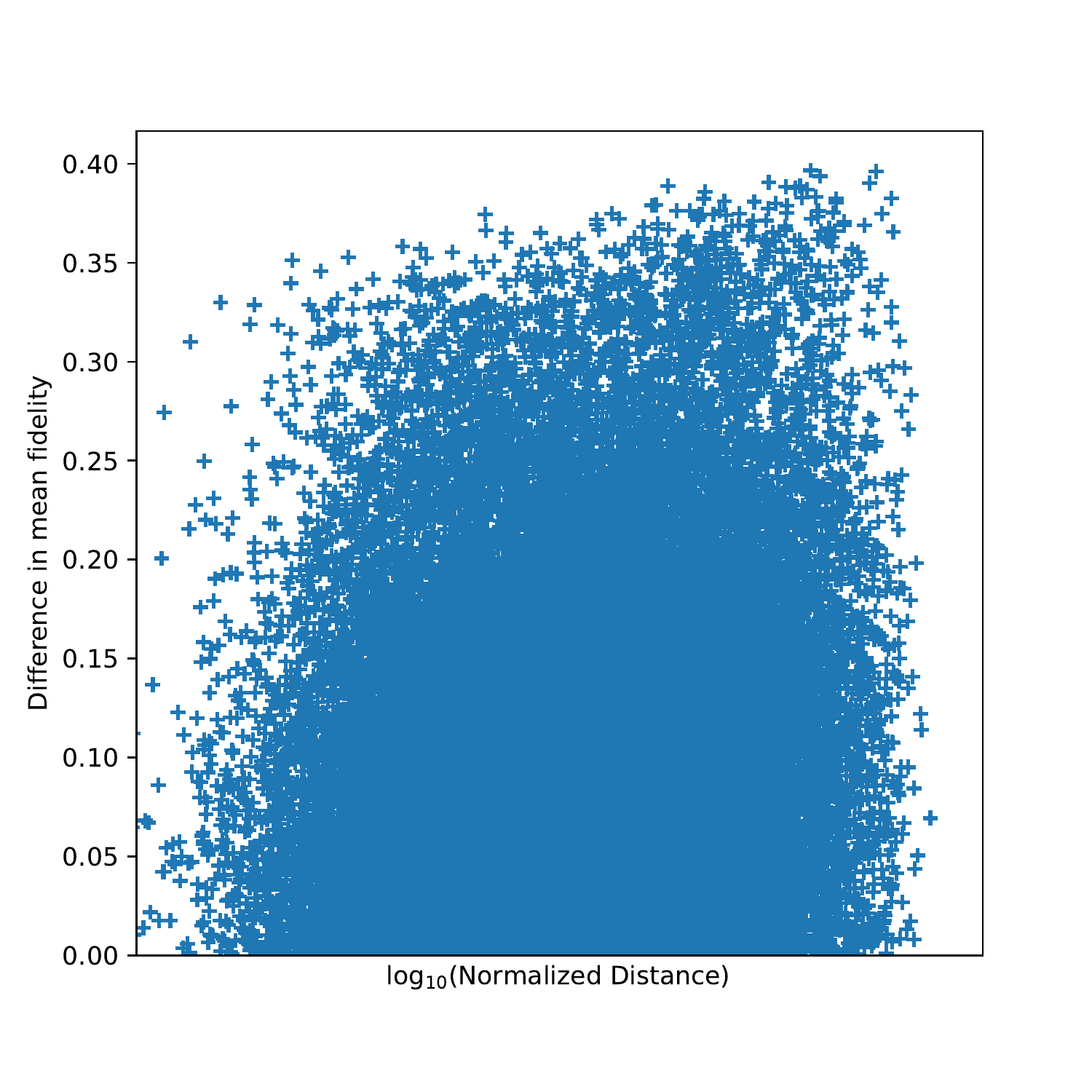}
    \end{subfigure}
    \caption{Each point in the scatter plot represents the distance and the performance after training of a pair of circuits.}
    \label{fig:dist-fid}
\end{figure}

Second, we demonstrate that the circuit metric aligns with the performance of QNNs in a simple training task. We examine 300 pairs of random QNNs with 3 qubits and a maximum of 20 gates. The QNNs are trained to mimic the action of a Quantum Fourier Transform (QFT) by maximizing the fidelity between the output states with ideal outcome states with the anchor states $\ket{\psi_k}$ as the input states, $\underset{\theta}{\max} \frac{1}{K} \|\bra{\psi_k} \text{QFT}^\dagger U(\theta) \ket{\psi_k}\|^2$. Fig.~\ref{fig:dist-fid} shows ~45000 pairwise distance and the difference in the circuits' performance. It can be seen that circuits with a small distance are likely to have the same learning capacity in terms of the optimal fidelity. However, it is inconclusive when the distance is large, since two good circuits, even with substantial architectural difference, could result in the same optimal fidelity.

\section{Implementation Details} \label{appdx:implementation}
\subsection{Optimizing the acquisition function}
We optimize the acquisition function \eqref{eq:EI} with an evolution approach. Beginning with a pool of $k(t)$ circuits, we generate $N_{\text{mut}}(t)$ samples by letting each initial circuit to reproduce $N_{\text{off}}(t)$ offsprings through mutation. We evaluate the acquisition function on $N_{\text{mut}}(t)$ samples. Those with high values are kept to the next generation. To be exact, half of the next generation are ones with highest values; the other half contains some of the remaining circuits such that those with higher values are more likely to be selected. Repeat the process through $\tau$ generations. We set $\tau = \Theta(\sqrt{t})$ and $k(t), N_{\text{off}}(t) = \Theta(\sqrt{\tau})$ by default. For each mutation, a circuit can change at most $4$ components. The number of changes from $1$ to $4$ is chosen from the distribution $[0.4,0.3,0.2,0.1]$. A change can be either replacing the gate at a random position or switching the qubit wires connected to a random gate.

\subsection{Training Quantum Neural Networks}
In all of our experiments, the black-box function involves optimizing quantum neural networks. The optimization problem is straightforward for QFT and MaxCut problems since the optimal value obtained from the training problem is also the black-box function. In particular, as discussed in Section \ref{Sec:Exp}, the objective function is $\max_{\theta} \frac{1}{K} |\bra{\psi_k} U_{\text{QFT}}^\dagger U(\theta) \ket{\psi_k}|^2$ for the QFT experiment and
$\frac{1}{M} \sum_{m=1}^M \max_{\theta_m} \frac{\bra{0} U^\dagger(\theta_m) H_m U(\theta_m) \ket{0}}{C_m}$ for the MaxCut experiment. Training quantum circuits for these tasks with the L-BFGS-B optimizer is fairly robust.

On the other hand, although we use the relative entropy $D_{KL}$ as the black-box function for BO, optimizing QGAN requires a different approach. Given $m$ training data $x^l$ and $m$ generated data $g^l$, the quantum generator $G_\theta$ and the classical discriminator $D_\phi$ are updated by $\min L_G$ and $\max L_D$ \cite{zoufal2019}, where
\begin{align}
    \nonumber L_G(\phi, \theta) =& \frac{1}{m} \sum_{l=1}^m \log D_{\phi} (g^l), \\
    L_D(\phi, \theta) =& \frac{1}{m} \sum_{l=1}^m [\log D_\phi (x^l) + \log(1-D_\phi (g^l))].
\end{align}
We train the QGAN objective using ADAM optimizer through $200$ epochs with batch size $100$ and learning rate $10^{-3}$. Parameters are initialized uniformly at random in the interval $(-0.1,0.1)$.

In a side note, we use a general version of the original MaxCut ansatz~\cite{farhi2014}, which applies only to unweighted graphs, For a weighted graph, the classical cost $C(x) = \sum_{i,j=1}^n w_{ij} x_i(1-x_j)$ can be mapped to $\sum_{(i,j)\in E} \frac{1}{2}w_{ij} (1-Z_i Z_j)$ with the mapping $x_i \mapsto \frac{1-Z_i}{2}$. Denote $H = \sum_{x \in \{0,1\}^n} C(x)\ket{x}\bra{x}$ the problem Hamiltonian and $H_B = \bigotimes_{i=1}^n \sigma^x_i$ the mixing Hamiltonian with Pauli $X$ operators. The MaxCut ansatz we use for experiments is $\ket{\psi(\bm{\alpha},\bm{\beta})} = U(\bm{\beta}_L) U(\bm{\alpha}_L) \dots U(\bm{\beta}_1) U(\bm{\alpha}_1) \ket{0}$, where $U(\bm{\alpha}_l) = \exp(-i\bm{\alpha}_l H)$ and $U(\bm{\beta}_l) = \exp(-i\bm{\beta}_l H_B)$. We optimize the parameter $\theta \equiv (\bm{\alpha}_l, \bm{\beta}_l)_{l=1}^L$ in the problem $\max_{\theta} \frac{\bra{0}U^\dagger(\theta)H U(\theta) \ket{0}}{C}$,
where $C \equiv \max_x C(x)$ is the true maximal cut. Since we have $M=10$ different graphs, there are $M$ slightly different MaxCut ansatzs, each applied to its corresponding graph. The blackbox function is the mean Hamiltonian value (normalized by the exact maximal cut) of every sub-problem.

\begin{figure}[h!] \centering
    \begin{subfigure}{0.49\textwidth} \centering
        \includegraphics[width=\textwidth]{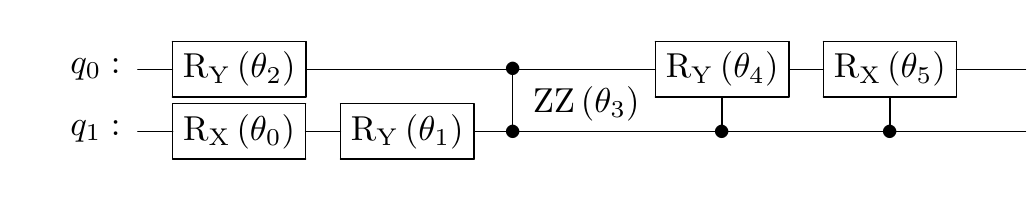}
        \caption{QFT}
    \end{subfigure}
    \begin{subfigure}{0.49\textwidth} \centering
        \includegraphics[width=\textwidth]{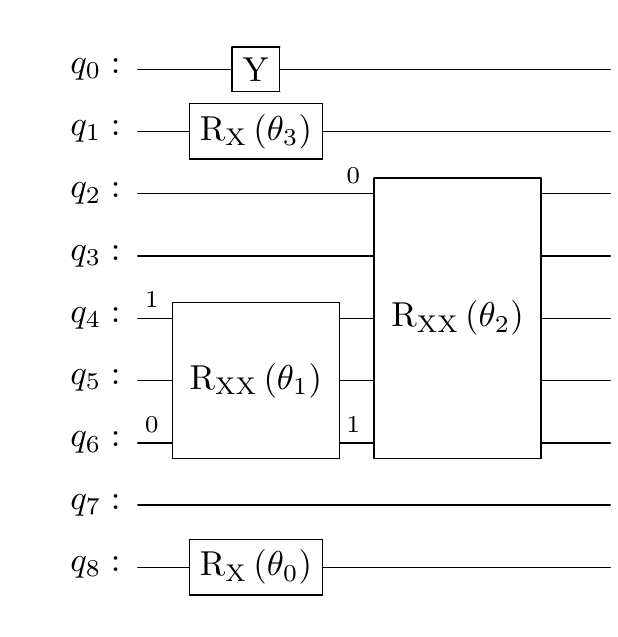}
        \caption{MaxCut}
    \end{subfigure}
        \begin{subfigure}{\textwidth} \centering
        \includegraphics[width=\textwidth]{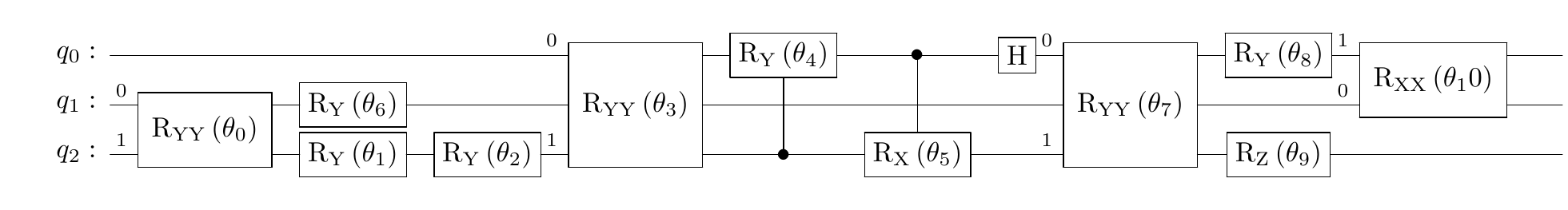}
        \caption{QGAN}
    \end{subfigure}
    \caption{Optimal circuits by BO for three learning tasks}
    \label{fig:best-circuits}
\end{figure}

\end{document}

%% file: topmatter.tex
\usepackage[utf8]{inputenc}
\usepackage{amsmath,amsfonts,amsthm,amssymb,physics}
\usepackage{graphicx}
\usepackage{url}
\usepackage{color}
\usepackage[usenames,dvipsnames,svgnames,table]{xcolor}
\usepackage[colorlinks=true, linkcolor=red, urlcolor=blue, citecolor=blue]{hyperref}
\usepackage[numbers]{natbib}
\usepackage{booktabs}
\usepackage{algorithm}
\usepackage{algorithmic}
\usepackage{bm}
\usepackage{stackrel}
\usepackage{enumitem}
\usepackage{breqn}
\usepackage{booktabs}
\usepackage{siunitx}
\usepackage{tabularx}

\usepackage[parfill]{parskip}

\newcommand{\Dc}{{\mathcal{D}}}
\newcommand{\Ec}{{\mathcal{E}}}

\newcommand{\Gc}{{\mathcal{G}}}
\newcommand{\Hc}{{\mathcal{H}}}

\newcommand{\Lc}{{\mathcal{L}}}
\newcommand{\Mc}{{\mathcal{M}}}
\newcommand{\Nc}{{\mathcal{N}}}

\newcommand{\Tc}{{\mathcal{T}}}
\newcommand{\Uc}{{\mathcal{U}}}

\newcommand{\Xc}{{\mathcal{X}}}

\newcommand{\Rbb}{{\mathbb{R}}}

\newtheorem{definition}{Definition}

\newtheorem{theorem}{Theorem}

\newcommand{\hilbert}{\Hc}
\newcommand{\R}{\mathbb{R}}

\newcommand{\txtstr}{\text{str}}

\newcommand{\txtgtm}{\text{gtm}}

\newcommand{\txtcore}{\text{core}}
\newcommand{\txtshape}{\text{shape}}

\newcommand{\txtlp}{\text{lp}}
\newcommand{\txtsp}{\text{sp}}
\newcommand{\txtavg}{\text{avg}}
\newcommand{\txtip}{\text{ip}}
\newcommand{\txtop}{\text{op}}

\newcommand{\Lone}{\Lc_1}
\newcommand{\Ltwo}{\Lc_2}

\newcommand{\bfone}{\mathbf{1}}

\usepackage{etoolbox}
\renewcommand{\bfseries}{\fontseries{b}\selectfont} 
\robustify\bfseries             
\newrobustcmd{\B}{\bfseries}